\documentstyle[seceq,wrapft,epsf]{ptptex}

\def\nuc#1#2{\mbox{${}^{#1}$\hbox{#2}}}
\newcommand{\nucl}[2]{\mbox{$^{#1}_{\scriptscriptstyle{\Lambda}}\hbox{#2}$}}
\newcommand{\nucL}[2] 
{\mbox{$^{#1}_{\phantom{1}\scriptscriptstyle{\Lambda}}\hbox{#2}$}}
\newcommand{\nucll}[2]{%
        \mbox{$^{#1}_{\scriptscriptstyle{\Lambda\Lambda}}\hbox{#2}$}}
\newcommand{\nucLL}[2]{%
   \mbox{$^{~\;{#1}}_{\scriptscriptstyle{\Lambda\Lambda}}\hbox{#2}$}}
\newcommand{\blank}{~\;}

\newcommand{\bold}[1]{\mbox{\boldmath $#1$}}    

\newcommand{\Z}{\bold{Z}}                       
\newcommand{\z}{\bold{z}}                       
\newcommand{\r}{\bold{r}}                       
\newcommand{\zc}{\bar{\bold{z}}}
\newcommand{\E}{{\scriptscriptstyle E}}
\newcommand{\Ezero}{{\scriptscriptstyle E_0}}

\newcommand{\Wei}{{\cal W}}
\newcommand{\VEV}[1]{\langle{#1}\rangle}        
\newcommand{\bra}[1]{\langle{#1}|}              
\newcommand{\ket}[1]{|{#1}\rangle}              
\newcommand{\F}{{\cal F}}                       
\newcommand{\SEV}[1]{\prec{#1}\succ}            
\newcommand{\Hml}{{\cal H}}                     
\newcommand{\Kin}{{\cal T}}	                
\newcommand{\Det}{{\rm det}}                    
\newcommand{\w}{{\bold{w}}}                     
\newcommand{\comment}[1]{}

\newcommand{\beq}{\begin{equation}}
\newcommand{\eeq}{\end{equation}}
\newcommand{\beqar}{\begin{eqnarray}}
\newcommand{\eeqar}{\end{eqnarray}}

\newcommand{\geteps}[2]{\epsfxsize=#1\epsfbox{PS/#2}}
\newcommand{\raf}[1]{(\ref{#1})}
\newcommand{\del}{\mbox{$\partial$}}
\newcommand{\lam}{\mbox{$\Lambda$}}
\newcommand{\alp}{\mbox{$\alpha$}}

\title{Quantum fluctuation effects on hyperfragment formation\\
from $\Xi^-$ absorption at rest on \nuc{12}{C}\footnote{%
This work was supported in part by
the Grant-in-Aid for Scientific Research
(Nos.\ 07640365, 08239104 and 09640329)
from the Ministry of Education, Science and Culture, Japan,
and by the Director, Office of Energy Research,
Office of High Energy and Nuclear Physics,
Nuclear Physics Division of the U.S. Department of Energy
under Contract No.\ DE-AC03-76SF00098.
}}

\author{
Yuichi    {\sc Hirata}$^{ a,}$\footnote{
               E-mail: hirata@nucl.sci.hokudai.ac.jp\ ,\ \ 
               Fax: +81-11-746-5444},
Yasushi   {\sc Nara}$^{ b}$, 
Akira     {\sc Ohnishi}$^{ a}$, 
Toru      {\sc Harada}$^{ c}$,
\\
and
J{\o}rgen {\sc Randrup}$^{ d}$
}

\inst{
$^a$	Division of Physics, Graduate School of Science,\\
        Hokkaido University, Sapporo 060-0810, Japan
\\
$^b$	
        Advanced Science Research Center,
        Japan Atomic Energy Research Institute,\\
        Tokai, Ibaraki, 319-11, Japan
\\
$^c$
        Department of Social Information,
        Sapporo Gakuin University,
        Ebetsu 069-8555, Japan
\\
$^d$
        Nuclear Science Division 70A-3307,
        Lawrence Berkeley National Laboratory,\\	
        Berkeley, California 94720, USA			
}

\recdate{
\today
}

\abst{
Formation mechanisms of single, twin, and double hypernuclei
from $\Xi^-$ absorption at rest on \nuc{12}{C} are investigated
with an refined microscopic transport model,
that incorporates the recently developed Quantal Langevin treatment
into Antisymmetrized Molecular Dynamics.
The quantum fluctuations suppress the formation probability
of double hyperfragments to around 10\%,
which is comparable to the experimental data,
and the dynamical formation of twin hyperfragment
can be described qualitatively.
}

\begin{document}

\maketitle

\section{Introduction}
\label{sec:intro}

%
%

%
%
Among various nuclear fragmentation processes,
the hyperfragment formation from the $\Xi^-$ absorption reaction at rest
is of primary importance for understanding strangeness in nuclei.
First, the $\Xi^-$ absorption is the most effective and the most
direct way to produce double $\Lambda$ nuclei.
For example, all three double hypernuclear formation events 
in which $\Delta B_{\Lambda \Lambda}$ is extracted
have been discovered through $\Xi^-$ absorption 
on light nuclear targets.\cite{DoubleA,DoubleB,DoubleC} \ 
Double hypernuclei give us valuable 
information of the low-energy $YY$ interaction,
and it might be a doorway to study multi-strangeness systems such 
as strange baryon matter,
which is expected to be realized in neutron stars.%
\cite{Multi}\ 
Because of this importance,
further experimental searches for 
double hypernuclear formation
through $\Xi^-$ absorption are being carried out and planned 
at BNL and KEK.\cite{BNL-E885,BNL-E906,KEK-E373} \ 
In these experiments, the number of stopped $\Xi^-$ is expected
to exceed the existing data by an order of magnitude.
Therefore, theoretical studies of the $\Xi^-$ absorption reaction
and the resulting fragmentations are urgently needed.
Second, in the KEK E176 experiment,
one finds interesting fragmentation patterns, 
such as \nuc{12}{C} + $\Xi^-$ $\rightarrow$ \nucl{4}{H} + \nucl{9}{Be}, 
in which two single hypernuclei are formed.\cite{TwinA,TwinB} \ 
This kind of reaction is called {\em twin} single hypernuclear formation.
It can be regarded as a fission of a $S=-2$ system
or an exotic decay, which is unexpected in light nuclei,
therefore it may suggest some special characteristics
of strangeness in nuclei.

In KEK E176,
the estimated number of  $\Xi^-$ absorption events at rest on 
light targets is $31.1 \pm 4.8$,
among which
one double hyperfragment formation event and two twin single hyperfragment
formation events were observed.\cite{Stopped}
In addition,
they found one event in which $S=-2$ hadronic system is formed	
(event type (B) in Table~\ref{table:Nakazawa}),
eight single hyperfragment formation events,
and eight other events, in which the visible energy release is greater than
28 MeV (event type (C) in Table~\ref{table:Nakazawa}). 
By using these experimental observations, 
the lower limits (90\% confidence level) 
of $S=-2$ and $S \leq -1$ sticking probabilities
were estimated as 4.8\% and 47.6\%, respectively.\cite{Stopped}\ 
In a similar way, we can estimate several lower and upper limits
of hyperfragment formation probabilities, as shown in Table~\ref{table:ULLimit}.

\begin{table}	
\caption{Observed stopped $\Xi^-$ events captured in light nucleus.
In events (A), the hyperfragment type (double, twin, or single) is specified.
In events (B), a hadronic system with $S=-2$ is formed (double or twin),
and at least one hypernuclear weak decay must occur in events (C)
($S=-1$ or $-2$).}
\label{table:Nakazawa}
\[
\begin{array}{llccc}
\hline
\hline
& \hbox{\ Hyperfragment} & \hbox{(A)} & \hbox{(B)} &\hbox{(C)}
\\
\hline
\begin{array}{l} S=-2 \\ \hbox{~} \\ S=-1 \end{array}
& \begin{array}{l} 
	\hbox{Double} \\ 
	\hbox{Twin} \\ 
	\hbox{Single}
	\end{array}
& \begin{array}{c} 1 \\ 2 \\ 8 \end{array}
& \begin{array}{c} \bigr\}\ +1 \\ ~ \end{array}
& \begin{array}{c} \Bigg\}\ +8 \end{array}
\\
\hline
\end{array}
\]
\end{table}

\begin{table}
\caption{%
Estimated lower and upper limit (90\% confidence level)
by using the number of events shown in Table \protect{\ref{table:Nakazawa}}.
As for the roughly estimated values, see the text.
}
\label{table:ULLimit}
\[
\begin{array}{cccccc}
\hline
\hline
\hbox{Hyperfragment} 
	&\multicolumn{3}{c}{\hbox{L. L. (\%)}}
	&\hbox{U. L. (\%)}
	&\hbox{Rough Est. (\%)}\\
\hline
\begin{array}{c} 
	\hbox{Double} \\ 
	\hbox{Twin} \\ 
	\hbox{Single}\\
	\hbox{No Hyp. Frag.}	\end{array}
& \!\!\begin{array}{c} - \\ 0.66 \\ 14.5 \\ - \end{array}
& \!\!\!\!\!\!\begin{array}{c} \bigr\}%
		\ 4.8\hbox{\protect\cite{Stopped}} \\ ~ \\ ~ \end{array}
& \!\!\!\!\!\!\!\!\!\!\begin{array}{c} \Bigg\}%
		\ 47.6\hbox{\protect\cite{Stopped}} \\ ~ \end{array}
& \!\!\!\!\begin{array}{c} 77.9 \\ 81.5 \\ - \\ 48.8 \end{array}
& \!\!\!\!\begin{array}{c} 3-9  \\ 6-18 \\ 26-73 \\ - \end{array}
\\
\hline
\end{array}
\]
\end{table}

These upper and lower limits are values which we must respect. 
However, the statistics is so poor that the constraints on the theory
is very loose at present.
Therefore, in this paper, 
we use very crude but plausible estimates of double, twin, and single
hyperfragment formation probabilities as follows.
One extreme would be to consider that
there were no more double and twin hyperfragment formation
other than the specified ones.
Then the double, twin, and single hyperfragment formation probabilities
can be estimated as 3\% (=1/31.1), 6\% (=2/31.1) and 26\% (=8/31.1), 
respectively.
Another extreme is to assume that
all the weak decay accompanied by charged particles are observed,
and the ratio among the double, twin and single hyperfragment formation
is kept.
Under this assumption, the probabilities might be estimated
as 9\% (= 1/11), 18\% (=2/11), and 73\% (=8/11), respectively. 
We should note that the probabilities of twin and double hyper fragment
formation are comparable.

%
%
Various analyses of the $\Xi^-$ absorption reaction have been carried out. 
These studies range from statistical approaches to direct reaction theories.
However,
no satisfactory consistent description of double and twin hyperfragment
formation is given yet.
%
%
With statistical decay models of the double hyperon compound
nucleus,\cite{Compound} \ 
the calculated double hyperfragment formation probability  
is too large as compared with that of twin hyperfragment formation,
provided that the $\Lambda\Lambda$ interaction is attractive.
Although the relative ratio of double- and twin-hyperfragment is improved
if the $\Lambda\Lambda$ interaction is assumed to be repulsive,
this assumption contradicts to previously observed double
hypernuclei.\cite{DoubleA}\ 
Thus the experimental data suggest the action of some dynamical effects
that cannot be mocked up by a simple escape probability of
one $\Lambda$ particle after the primary elementary reaction,
$\Xi^- p \to \Lambda\Lambda$.
%
%
%
On the other hand, 
Yamada and Ikeda have proposed an intuitive 
direct fragmentation picture which is based on the symmetry of
the target ($\nuc{12}{C}$) wave function:\cite{Yamada,YI97} \ 
When proton $s$-hole and $p$-hole states are created in the 
primary elementary reaction and a highly excited double hypernuclear
state is formed, it is fragmented mainly at the doorway stage
to various hypernuclei by emitting some nucleons or clusters. 
Following this picture, they obtain that the excited
channels of \nucll{12}{Be} + p, \nucll{12}{B} + n, and
\nucll{11}{Be} + d are produced more strongly than other channels,
and the calculated $S=-2$ sticking probability is
in agreement with the KEK E176 experiment.
However, this model still underestimates the twin hypernuclei 
formation probability.\cite{YI97} \ 

%
%
The aim of this paper is to investigate the formation mechanism
of single, twin, and double hypernuclei 
from the $\Xi^-$ absorption at rest on \nuc{12}{C}
by applying a recently developed microscopic transport model
that augments the Antisymmetrized Molecular Dynamics (AMD)
model~\cite{AMD}
by the effect of quantum fluctuations as given by the
Quantal Langevin model.\cite{OR95,OR97a,OR97b} \ 

%
%
The microscopic transport approach can describe a variety of processes,
from fast reactions, such as quasi free processes that are mainly determined
by the elementary two-body collisions,
to more central reactions where bulk collective motion occurs
and the mean-field dynamics is important.
In addition, combined with statistical decay models, 
the light-particle evaporation from moderately excited nuclei can be described.
Since we want to extract the contribution of dynamical processes to
the hyperfragment formation from the $\Xi^-$ absorption,
the applicability to both dynamical and statistical processes
is an important advantage of the transport approach.

%
%
Among various microscopic transport models,
AMD and Fermionic Molecular Dynamics (FMD)~\cite{FMD}
take account of the fermionic nature of baryons
by using anti-symmetrized wave packets.
This feature makes it possible to describe
fragmentation processes due to cluster and nuclear 
shell effects,\cite{AMD,FMD,Cool1,Cool2} \ 
which are important for the study of light nuclear system, 
including hypernuclei.\cite{Lam-rang} \ 
Especially two-body collision processes, 
which are indispensable for the description of nuclear reactions,
are incorporated into AMD.
Because of these advantages,
we adopt AMD combined with statistical decay model as the 
starting point for studying $\Xi^-$ absorption reactions.
%

%
%
However, even if anti-symmetrized wave packets are employed,
the use of product wave functions,
whether of Gaussian form or of the Hartree-Fock type,
has certain generic shortcomings that may be of specific importance
in the present context.
%
%
In particular, it is usually difficult to describe
the dynamical formation of 
ground-state fragments in the outgoing state.
This point was already noted and some remedies
were proposed.\cite{AMD,Danielewicz,NOH95} \ 
In the hyperfragment formation problem, for example, 
Nara et al. suggested the necessity of the  
direct productions $K^- + \alpha \to \pi^0 + \nucl{4}{H}$
in the $K^-$ absorption reaction at rest on clusterized light nuclear targets
which cannot be described with only two-body collision terms.\cite{NOH95} \ 

%
%
Another possibly important problem of microscopic transport models
concerns the inherent fluctuations associated with wave packets.
When the functional space of single-particle wave functions is 
restricted to Gaussian form and the system is described 
by a single Slater determinant,
the neglected degrees of freedom
are expected to cause fluctuations of the retained degrees of freedom.
%
%
A treatment of this inherent problem has been proposed recently
by Ohnishi and Randrup
in the form of the Quantal Langevin model.\cite{OR95,OR97a,OR97b}\ 
With the development of this model,
they have discussed the importance of the fact that
wave-packet wave functions are not eigenstates of the Hamiltonian operator
and thus possess inherent energy fluctuations
that affect both the statistical properties
and fragmentation reactions in heavy-ion collisions.

%
%
The above features may affect the fragmentation	
from the $\Xi^-$ absorption reaction.
First, the observed twin hyperfragments are in 
their ground states or low excited states.
Next, if the excitation energy is shared among all the degrees of 
freedom, each particle will have a smaller energy than the separation energy
and then particle evaporation and fragmentation may be artificially suppressed
in molecular dynamics calculations.
For example, in the $\Xi^-$ absorption reaction, 
many large excited fragments like 
\nucll{13}{B} and \nucL{12}{B} are  produced in the dynamical 
simulation of AMD as discussed later.
This consideration imposes the importance of the quantum statistical
features,
which enhances ground-state fragments since 
internal degrees of freedom of fragments are likely to be {\em frozen}
while the relative motions between fragments are agitated.

Therefore, in this paper, we focus our attention on 
the effects of the quantum fluctuations.
%
%
Specifically,
we extend the usual AMD transport model
to incorporate quantum energy fluctuations
in the manner of the Quantal Langevin model.
We call the resulting model AMD-QL.
%
%
We will show that the application of AMD-QL to $\Xi^-$ absorption at rest
on $\nuc{12}{C}$ leads to total formation probabilities of various double 
hypernuclei of about 10\%, 
when the $\Lambda\Lambda$ interaction is attractive.

%
%
This paper is organized as follows:
We first briefly describe
how we can incorporate quantum fluctuations in wave packet dynamics 
in the manner of the Quantal Langevin model in Sec.~\ref{sec:QL}.
In Sec.~\ref{sec:AMD-QL}, we incorporate this quantum fluctuation
into AMD, and give the form and the strength of the fluctuation.
Then we show our results of hyperfragment formation in $\Xi^-$
absorption at rest in Sec.~\ref{sec:results}.
In Sec.~\ref{sec:sum}, we summarize our work.

\newpage

\section{Quantum fluctuations in wave packet dynamics}\label{sec:QL}

As already mentioned in the introduction, 
the fluctuations are not fully incorporated
in the transport models describing
the time evolution of wave packets
based on the time-dependent variational principle (TDVP).
In a statistical context, 
this was already pointed out in Refs.~\citen{OR95,OR97a,OR97b,OR93}. 
There are some claims that the statistical properties 
of the wave packet dynamics can be interpreted to be quantal
rather than classical
by re-scaling the temperature to fit the exact excitation energies
and by analyzing the wave function itself through, for example, 
the single-particle energy spectrum~\cite{SF96,OH96b}.
However, 
this kind of interpretation does not work in the case of fragment formation,
since the fragments are described by the centroid parameters of 
nucleon wave packets.
Thus the same authors of Refs.~\cite{SF96,OH96b} had to introduce
additional fluctuations or correlations to describe fragmentation
processes.\cite{OH96b,OH96a,SF97,OH96c}\ 

Up to now, several origins of fluctuation in wave packet dynamics
have been proposed,
such as the momentum width in a single-particle wave packet,\cite{OH96a}\ 
short-range correlation between nucleons,\cite{SF97}\  
or the lack of self-consistency between the wave packet 
and the mean field.\cite{OH96c}\ 
In this work, we do not specify the source of fluctuations,
but instead require that the quantum statistical equilibrium is approached
in the course of the time evolution
following the Quantal Langevin model.\cite{OR95,OR97a,OR97b}
This approach has the merit that the equilibrium properties are ensured.
On the other hand,
since the system approaches the same equilibrium
with any fluctuations within the Quantal Langevin model,
we have to determine the form and strength of the fluctuation
separately to describe dynamics.
In this paper,
we parametrize the fluctuation in a relatively simple form 
and seek to determine the allowed range of the fluctuation strength
by applying the model to some phenomena with dissipation and/or fluctuation.

In this section, we briefly describe the Quantal Langevin model
and its application to dissipative nuclear collective motion,
simulated within a simple soluble model, the Lipkin model.\cite{Lipkin}

\subsection{Quantal Langevin Model}
\label{subsec:QL}

The Quantal Langevin model is designed to ensure
that the equilibrium properties of the system
are in accordance with quantum statistics.
In the case of canonical and microcanonical ensembles,
the statistical properties are governed by 
the partition function and the microcanonical phase volume, respectively,
\beqar
\nonumber
{\cal Z}(\beta) = {\rm Tr}\left( \exp( - \beta \hat H) \right)
        = \int d\Gamma\, \Wei_\beta(\Z)\ , &&\quad
\Wei_\beta(\Z) = \VEV{\Z|\exp(-\beta \hat H) |\Z}\ ,\\
&& \label{eq:canonical}
\\
\nonumber
\Omega(E) = {\rm Tr}\left( \delta (E - \hat H) \right)
        = \int d\Gamma\, \Wei_\E(\Z)\ , &&\quad
\Wei_\E(\Z) = \VEV{\Z|\delta(E - \hat H) |\Z}\ ,
\\
&& \label{eq:microcan}
\eeqar
where $\ket{\Z}$ represents a parametrized and normalized quantum state,
and $\int d\Gamma\, \ket{\Z}\bra{\Z}$ resolves unity. 
From these expressions, 
we see that the probability to find a state $\ket{\Z}$ is proportional to 
the statistical weight $\Wei(\Z)= \Wei_\beta(\Z)$ or $\Wei_\E(\Z)$.

%
%
To produce the desired equilibrium distribution
$\phi(\Z;t) \propto \Wei(\Z) \equiv \exp(-\F(\Z))$,
it is possible to adopt the fluctuation-dissipation dynamics
described by, for example, the Fokker-Planck equation, 
\beqar
\label{eq:FP}
{D \phi(\Z;t) \over Dt}\
        &\equiv&\ {\del \phi \over \del t} +\ \{ \phi , \Hml \}_{P.B.}
	=\ 
	-\sum_i \frac{\del}{\del q_i}
                \left( V_i -\ \sum_j M_{ij} \frac{\del}{\del q_j} \,
                \right) 
		\phi\ ,
        \\
V_i\    &=&\ -\sum_j\, M_{ij}\, \frac{\del \F(\Z)}{\del q_j}\ ,
\eeqar
where $\{q_i\}$ are canonical variables satisfying $d\Gamma = \prod_i dq_i$,
$D/Dt$ represents the time derivative along the classical path,
and $\Hml$ denotes the classical Hamiltonian.
The second relation
is recognized as the Einstein relation that follows
from the requirement that the equilibrium distribution 
be a static solution of Eq.~(\ref{eq:FP}).
In numerical simulations,
it is easier to treat
the corresponding Langevin equation, 
\beq
\label{eq:LangevinE}
\frac{D q_i}{Dt}\ 
        \equiv\ \dot{q}_i -\ \{ q_i, \Hml \}_{P.B.}
	= V_i\ +\ \sum_j g_{ij} \zeta_j \ ,
\eeq
where the fluctuation matrix $\bold{g}$ is related
to the mobility tensor $\bold{M}$
through $\bold{g}\cdot\bold{g} = \bold{M}$ 
and $\zeta(t)$ is the white noise, 
$\SEV{\zeta^*_i(t)\, \zeta_j(t')} = 2\delta_{ij}\,\delta(t-t')$.
Here we have ignored the diffusion-induced drift term.\cite{Risken} \ 
The R.H.S. of Eq.~(\ref{eq:LangevinE}),
which includes both the average drift term and the stochastic diffusion term
described with the white noise $\zeta$, 
is referred to as the {\em Quantal Langevin} force
and it gives the energy fluctuation for the system. 
This random force arises from the energy dispersion of each wave packet
and thus has a purely quantal origin.
Numerically Eq.~(\ref{eq:LangevinE}) can be solved by
the finite difference method in which white noise integrated over a short
period $\Delta t$ is treated as
a random number following the complex normal distribution
generated independently in each time step,
multiplied by $\sqrt{2\Delta t}$. 

In a practical application, we use the ansatz for the statistical
weight $\Wei(\Z)$ by applying the harmonic approximation.\cite{OR97a}\ 
For example, the statistical weight in a microcanonical ensemble
$\Wei_\E(\Z)$ is assumed to be a continuous Poisson distribution,
\beqar
\label{MCprobPoi}
\Wei_\E(\Z)\ 
	&\propto&\   e^{-\Hml/D}\ {(\Hml/D)^{E/D} \over (E/D)!}\
        =\      e^{-\Hml/D}\ {(\Hml/D)^{E/D} \over \Gamma(E/D+1)}\ ,\\
D
	&=& {\sigma^2_\E \over \Hml}\ , \quad
\sigma_\E^2
	= \VEV{\Z|\hat H^2|\Z}\ - \VEV{\Z|\hat H|\Z}^2\ ,
\eeqar
where the ground-state energy is subtracted from the energy expectation
value $\Hml$.
Here, $D$ denotes a typical energy scale.
Since this typical energy scale $D$ depends on $\Z$ only weakly
in many cases, the drift term is simplified as follows,
\beq
V_i\   =\ -\beta_\Hml\,\sum_j\, M_{ij}\, \frac{\del \Hml}{\del q_j}\ ,
\quad
\beta_\Hml\ =\ {\Hml - E \over \sigma^2_\E}\ ,
\eeq
where $\beta_\Hml$ denotes a state dependent inverse temperature
modified by quantum correction.\cite{OR97a}\ 
With this form of the drift term, the Einstein relation becomes
manifest. In addition, once the energy dispersion 
$\sigma^2_\E$ is given, it becomes feasible to solve the Quantal Langevin
equation without further complication.


In addition to the appearance of a random force in the equation of motion,
we have to take the intrinsic distortion of wave packets into account
before making any observation by using the wave packet ensemble.
This point can be easily understood by considering the definition of
statistical mean value of an observable $\hat O$ in canonical ensemble.
\beq
\SEV{\hat O} 
	= {1\over {\cal Z}} {\rm Tr}\left(\hat O\,e^{-\beta\hat H} \right)
	= {1\over {\cal Z}}
        \int\, d\Gamma\, \Wei_\beta(\Z)\,
		{
		\VEV{\Z|e^{-\beta\hat H/2}\,\hat O\,e^{-\beta\hat H/2}|\Z}
			\over
		\VEV{\Z|e^{-\beta\hat H}|\Z}}
		\ .
\eeq
Since the Boltzmann weight operator $\exp(-\beta\hat H)$ cannot be
treated as a $c$-number,
the distorted wave function $\ket{\Z'}=\exp(-\beta\hat H/2)\ket{\Z}$
is different from the original evolving state, $\ket{\Z}$.
For example, the distortion operator $\exp(-\beta\hat H/2)$ emphasizes
the energy eigencomponent with smaller energies 
and therefore the observed energy
$\Hml_\beta=\VEV{\Z'|\hat H|\Z'}/\VEV{\Z'|\Z'}$ is smaller than
$\Hml=\VEV{\Z|\hat H|\Z}/\VEV{\Z|\Z}$.\cite{OR97a,OR97b,OR97c}\ 

\begin{wrapfigure}{l}{\halftext}
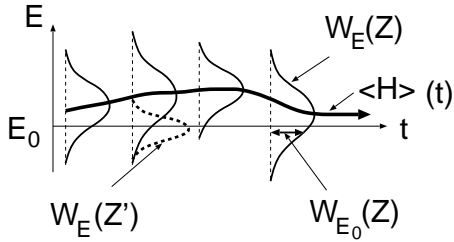

\begin{center}
\geteps{6cm}{Eflct.eps}
\end{center}
\caption{%
Fluctuation of the energy expectation value.
In the Quantal Langevin treatment with a specified energy $E_0$,
the expectation value with respect to the evolving state $\ket{\Z}$
fluctuates
(if that wave packet has a component of the eigenstate of the energy $E_0$).
However, any observation involves the distorted state
$\ket{\Z'}=\exp(-\beta\hat H/2)\ket{\Z}$.
}
\label{fig:Eflct}
\end{wrapfigure}
When the total energy $E_0$ is specified,
the wave packets are distributed according to the weight $\Wei_\Ezero(\Z)$
in the Quantal Langevin model, 
as is shown schematically in Fig.~\ref{fig:Eflct}. 
Thus the expectation value of the Hamilton operator
with the evolving state $\ket{\Z}$ can fluctuate around the specified energy.
The difference between the evolving state $\ket{\Z}$
and the distorted state $\ket{\Z'}$
explains why this fluctuation of energy is allowed:
When we make any observation,
we have to use expectation values with the distorted state
$\ket{\Z'}=\sqrt{\delta(E_0-\Hml)}\ket{\Z}$,
which is an energy eigenstate with the eigenvalue $E_0$.
Therefore, there is no fluctuation in the observed energy,
while the fluctuation of the energy expectation value
with respect to the evolving state is necessary.
We will discuss this point in the next subsection.

\subsection{Application to Lipkin Model}
\label{subsec:Lipkin}

In equilibrium, the Quantal Langevin model has been shown to work 
very well and give reasonable statistical properties, including 
the quantum statistical features of simple soluble systems
and nuclei.\cite{OR95,OR97a,OR97b,OR93}\ 
In a dynamical context,
the fragmentation processes in heavy-ion collisions
have been studied.\cite{OR97b}\ 
However,
since exact dynamical results are not available,
the model has not been verified as a general theory
to describe fluctuation-dissipation dynamics.
%
In order to examine this aspect,
it may be instructive to employ a schematic model,
such as the Lipkin model,\cite{Lipkin}\ 
to study the damping of collective motion,
since this is one of the most familiar dissipative phenomena
in nuclear physics.
%

The Hamiltonian operator of the Lipkin model,
\beq
H
	= \epsilon\, K_0 - {1\over 2}\,V\,
	\left(  K_+\, K_+ +  K_-\, K_- \right)\ ,\\
\eeq
contains only the quasi-spin operators,
\beq
K_0
	\equiv {1\over 2}\sum_{n=1}^N\, 
	\left( c^\dagger_{+m}\,c_{+m} - c^\dagger_{-m}\,c_{-m} \right) \ ,\quad
K_+
	\equiv \sum_{n=1}^N\, c^\dagger_{+m}\,c_{-m}\ ,\quad
K_-	\equiv	(K_+)^\dagger\ .
\eeq
Therefore, the collective subspace that couples to the unperturbed
ground state $\ket{0}$ completely decouples from other subspaces.
Within this collective subspace, the wave packet
specified by the complex parameter $z$,
\beq
\ket{z} \equiv \exp(zK_+)\ket{0}\ ,
\eeq
describes a general product wave function which couples to $\ket{0}$.
We study the damping of collective motion through the time evolution
of this wave packet by using the following Quantal Langevin equation
for the canonical variable $w=z\,\sqrt{N/(1+\bar{z}z)}$, 
\beqar
\label{eq:QL-Lipkin}
{Dw\over Dt}
	&=& \dot{w}
	- {1\over i\hbar}\,{\del\Hml\over\del\bar{w}}
	= 
		- \beta_\Hml\,g^2\,{\del\Hml\over\del\bar{w}}
		+ g\,\zeta\ ,\\
g^2	&=& {g_0^2 \sigma_\E \over \hbar}\ ,
\quad
\sigma_\E^2
	= {\del\Hml\over\del z}\,C^{-1}\,{\del\Hml\over\del\bar{z}}\ ,
\quad
C	= {N \over (1+\bar{z}z)^2}\ .
\eeqar
When we solve this equation in $z$ space,
there naturally appears an $N$ dependence in the coefficient,
$g^2_z \simeq g^2\,\sigma_\E/N\hbar$, in the lowest order of $z$,
which comes from the transformation between $z$ and $w$.
We have checked that the qualitative features of the results shown below
do not depend on the detailed structure of this coefficient $g_z^2$,
such as its small-$z$ dependence,
if this $N$ dependence is preserved.

\begin{figure}
\centerline{~\hspace*{3cm}~\geteps{10cm}{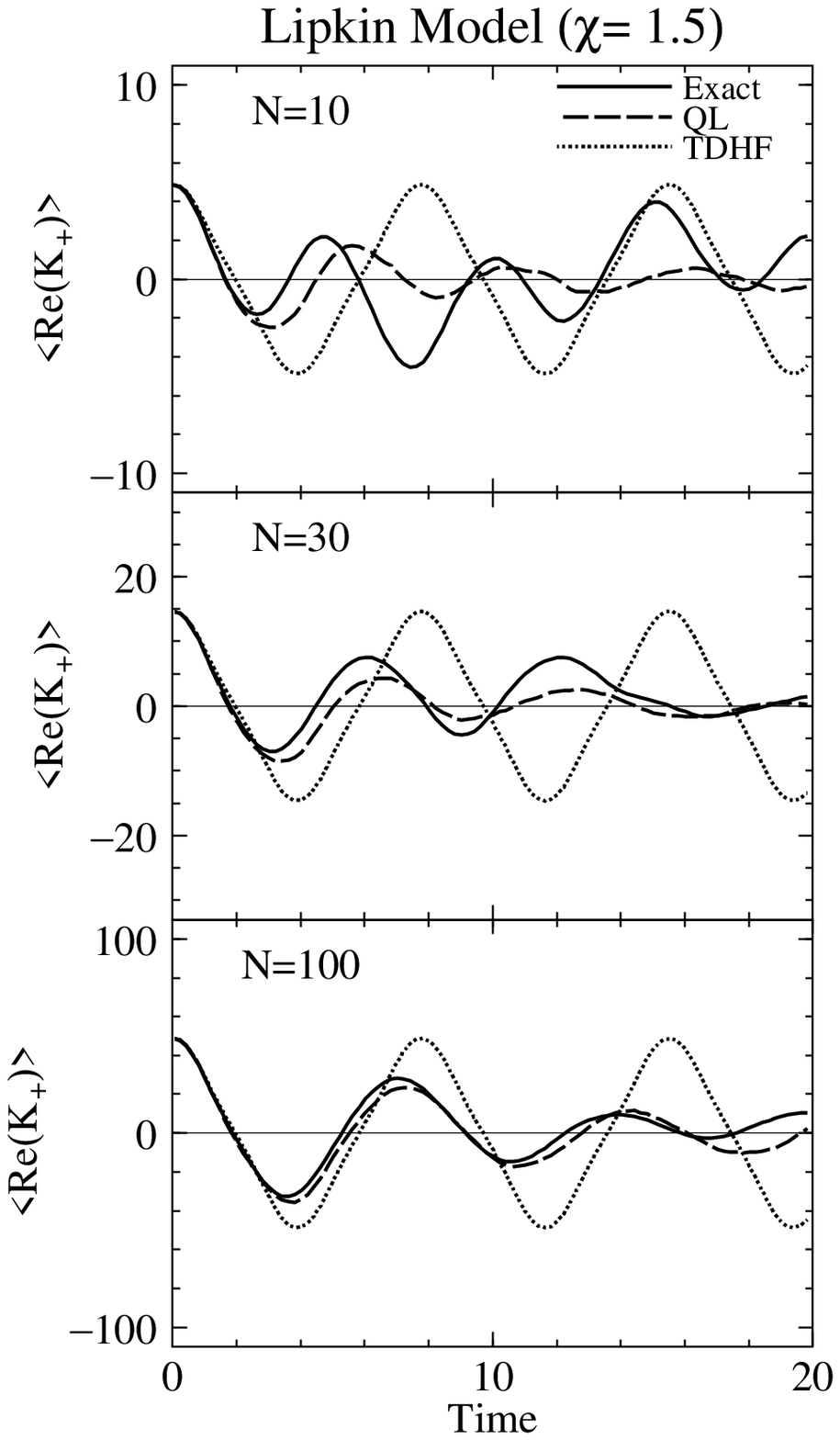}~\hspace*{-4cm}%
~\geteps{10cm}{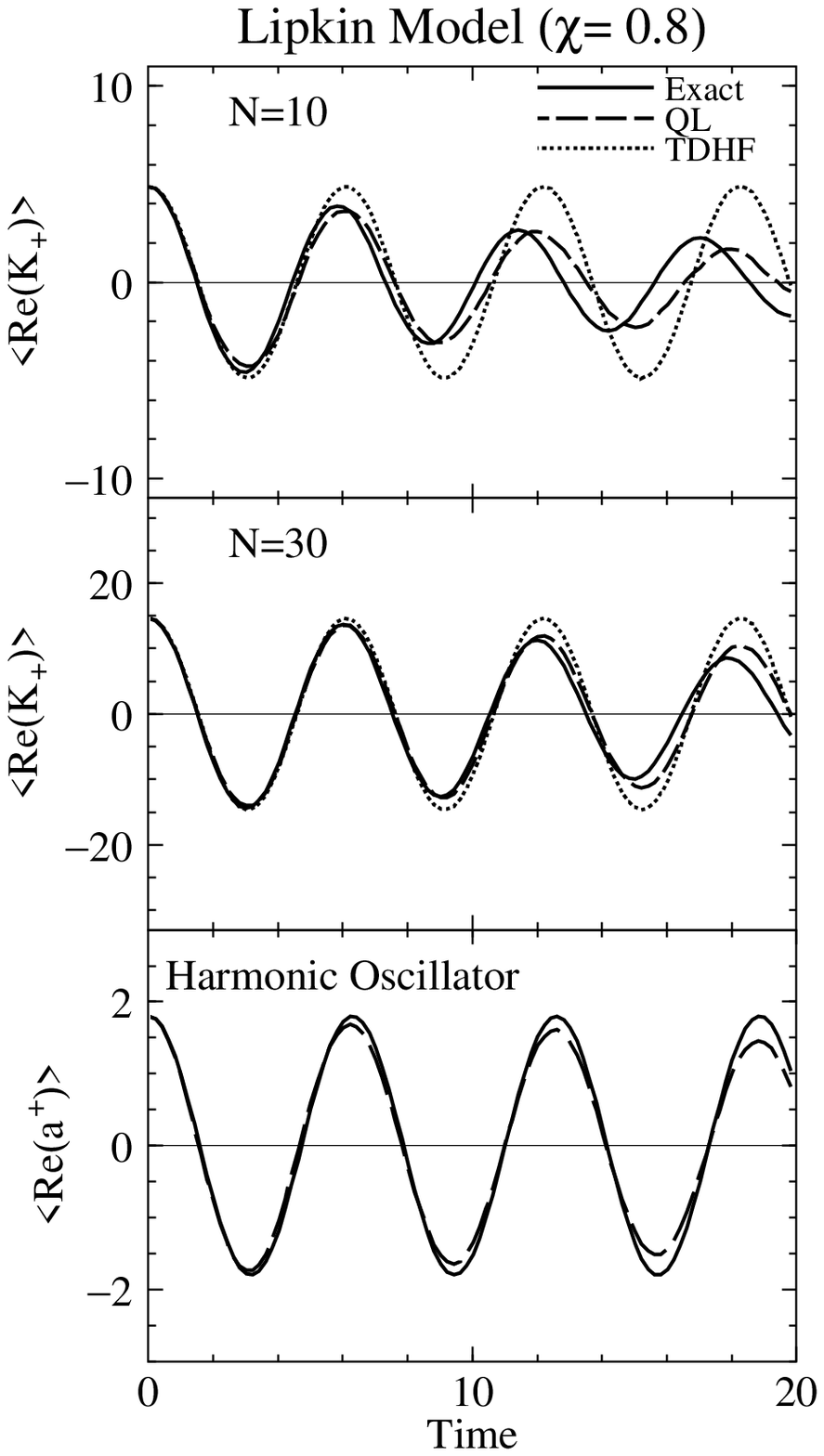}~}
\vspace*{-4cm}
\caption{
Damping of collective motion in the Lipkin model.
Solid, dotted, and dashed curves show the exact solution
and the results of TDHF and the Quantal Langevin model, respectively.
The left (right) panels show the results with
$\chi=V (N-1)/\epsilon = 1.5$ (0.8),
leading to a double (single) potential well.
For comparison the bottom-right panel presents the case of
the pure harmonic oscillator
(one Gaussian wave packet in a harmonic oscillator potential).
}
\label{fig:Lipkin}
\end{figure}

By using the Langevin Eq.~(\ref{eq:QL-Lipkin}), 
we have generated the ensemble of evolving paths.
At each time, we have calculated the ensemble average
of the operator $\hat K_+$ with the distorted wave function, 
while keeping the evolving state as it is.
The strength of the potential is chosen to keep
the average potential effect constant. Namely, the following
$\chi$ parameter is chosen to be independent on $N$,
\beq
\chi \equiv {V\over \epsilon}\,\left( N - 1\right)\ .
\eeq
When this parameter is less than unity,
the potential energy surface has only one well,
while it becomes a double-well shape for $\chi > 1$.
As typical examples, we show the calculated results
with $\chi=1.5$ and 0.8
and $N$=10, 30, and 100,
and the initial condition $z(t=0)=0.6$.
As for the fluctuation strength parameter, 
we show the results with $g_0$=0.3. 
However, the qualitative behavior does not change 
in the wide range of this parameter, $0.1 \leq g_0 \leq 0.5$.

%
%
In Fig.~\ref{fig:Lipkin}, we show the time development
of $\langle{\rm Re}K_+\rangle$, which is related to the collective coordinate.
For comparison we also show the exact solution and the results of TDVP
by using the above product type wave function.
Since the above wave function describes a general product wave function,
the results of TDVP is nothing but that of the time-dependent
Hartree-Fock (TDHF) method.
The exact solution shows interesting behavior.
At $\chi$=1.5, the collective motion gradually damps at larger $N$,
while for $N=10$ the collective variable becomes larger again
owing to the interference between some specific energy eigencomponents.
In addition, we can see some hardening of the collective mode
at smaller $N$.
For example, one period becomes much smaller than the TDHF results.

The Quantal Langevin model results exhibit a behavior similar
to the exact solution, although 
it is impossible to treat the interference between some specific
energy eigencomponents which agitates the collective motion again
after several periods.
The average damping width seems to be well reproduced,
and more than half of the hardening of this mode is explained.
On the other hand, the TDHF result shows a pure periodic motion,
since there is only one complex degree of freedom in this wave packet.
If the potential strength parameter $\chi$ is kept constant,
there is no room for $N$ to play any role in TDHF.
Therefore, the results of TDHF shows the limit of $N\to\infty$ 
without any damping.

Within the Quantal Langevin model,
the above mentioned damping and hardening 
can be explained as follows:
Each wave packet $\ket{z}$ has its inherent energy fluctuation
$\sigma_\E$ which is proportional to $\sqrt{N}$,
and in the Quantal Langevin treatment,
the energy expectation value of this evolving state
can fluctuate around the specified energy 
with the order of this intrinsic energy fluctuation.
Therefore, each wave packet will have different energy expectation value
and frequencies and, as a result of ensemble average, 
the collective motion damps. 
The hardening of the collective mode is also explained similarly.
In the case of $\chi$=1.5 there is a barrier in the potential 
energy surface at ${\rm Re}(z)$=0 and the energy of the initial state
is just above this barrier height.
In the Quantal Langevin treatment,
some of the evolving paths are trapped by this barrier
and return to the neighborhood of the initial state
much earlier than the classical motion.
The first peak of the collective variable for $N=10$
comes from these evolving states.
These features may be related to the 
strange behavior of collective motion at around the bifurcation
point.

When $N$ becomes larger, the importance of the 
above intrinsic energy fluctuation $\sigma_\E \propto \sqrt{N}$
relative to the specified energy $E \propto N$ becomes smaller.
Therefore the damping and the hardening of mode become smaller.

In the semi-harmonic case $\chi=0.8$ 
the situation is similar.
The Quantal Langevin model describes the damping well,
but since the potential energy surface is a single well
and there is no trapping by the barrier,
the behavior of the collective variable is very regular and the damping
width becomes smaller.
As a limit, we can consider pure harmonic motion.
For example, we show in the right-bottom panel of Fig.~\ref{fig:Lipkin}
the time evolution of $\langle a^\dagger\rangle$ in the case of one wave packet
in a harmonic oscillator potential.
In this case, TDHF (equivalent to AMD in this case) gives the 
exact solution of the time-dependent Schr\"odinger equation,
while the Quantal Langevin model still gives a small damping.
However, this small damping occurs in a special situation
and becomes visible only after many periods.
Thus it is not a big problem
since we are interested in the bulk behavior of dissipative systems
such as nuclei.

\section{Antisymmetrized Molecular Dynamics with Quantal Langevin Force}
\label{sec:AMD-QL}

%
\subsection{Implementation of the Quantal Langevin Force into AMD}
\label{subsec:AMD-QL}

%
%
In AMD, the quantum states are constructed by the following 
Slater determinant of Gaussian wave packets,\cite{AMD} \ 
\beqar
\label{Twave}
\ket{\Z}
        &=&     {1\over \sqrt{A!\ \Det \bold{B}}}\
                \Det \bigl( \ket{\z_i(\bold{r}_j)} \bigr) \ , \\
\label{Swave}
\ket{\z_i(\bold{r}_j)}
        &=&     \left({2\nu_i\over\pi}\right)^{3/4}
                \exp\left[
                        -\nu_i(\r_j-\z_i/\sqrt{\nu_i})^2+{1\over 2}\z_i^2
                \right] \chi_i(j) \ , 
                \\ 
B_{ij}	&=&	\VEV{\z_i|\z_j}\ ,
\eeqar
   where $\chi_i$ represents the spin-isospin wave function and 
   the parameter $\nu_i$ is (inversely) related to the variance
   of the Gaussian wave packet;
   both are assumed to remain constant in time.
For nucleons we use $\nu_N=0.16$ fm$^{-2}$. 
%
%
We have included $\Lambda$ and $\Xi^-$ particles in the framework of AMD
and
   the width parameters for these particles are chosen to be 
   $\nu_Y=(m_Y/m_N)\nu_N$, where $m_Y$ and $m_N$ are masses of 
the hyperon and nucleon, respectively.
This prescription makes it possible to factorize the center-of-mass 
wave function. 
%
%
The real and imaginary parts of the parameter $\{ \z_i \}$ of the
   Gaussian wave packet (\ref{Swave}),
\beq
\label{z}
        \z_i=\sqrt{\nu_i}\bold{d}_i + {i\over2\hbar\sqrt{\nu_i}}\bold{k}_i \ ,
\eeq
represent the mean position $\bold{d}_i$ and 
the mean momentum $\bold{k}_i$, respectively.
%
%
Applying the time-dependent variational principle
to total wave function $\ket{\Z}$,
\beq
\delta  \int dt\ \VEV{\Z| i \hbar \frac{ \del }{ \del t} - \hat{H} |\Z} = 0\ ,
\eeq
we obtain the equation of motion for the parameters $\{ \z_i \}$,
\beqar
\label{eom}
{D\z_i \over Dt} &\equiv&
        \dot{\z}_i - \frac{i}{\hbar} \bold{F_i} = 0\ ,\\
\bold{F}_i &=&  -\sum_{j} C^{-1}_{ij} {\partial\Hml \over \partial \zc_j}\ ,\\
C_{ij}   &=& 
\frac{ \del^2 }{ \del \zc_i \del \z_j } \log { \det \bold{B} }\ .
\eeqar
Here, 
$\Hml= \VEV{\Z | \hat{H} | \Z}$
is the expectation value of the total energy.

%
%
We can construct the Quantal Langevin equation based on the AMD wave
functions as before,
although there are several points that require special consideration.
%
%
The first problem is the treatment of the zero-point CM kinetic energies
of fragments.\cite{AMD,FMD,Kiderlen}\ 
Since this fragment zero-point CM kinetic energy is proportional
to the number of fragments and does not disappear even in the asymptotic region,
it modifies the $Q$ value of fragmentation 
and suppress fragmentation artificially.
This fragment zero-point CM motion also affects the energy dispersion.
For example, when one isolated fragment moves
the energy fluctuation grows as a linear function of the translational
kinetic energy of this fragment.
In wave packet dynamics with product-type wave functions,
where the fragment CM motion is described by
a wave packet, these zero-point kinetic energies are indispensable.
However, if we consider a superposition of the wave packets
with respect to impact parameter and time
in order that the incident wave function
be a plane wave as in actual reactions,
then the relative phases of this superposition are kept in time
and the zero-point CM kinetic energies of the fragments will disappear
 in the asymptotic region.
Therefore, the above effects are spurious
and should be removed from the Hamiltonian
as well as from the energy dispersion.
Several prescriptions for subtracting
fragment zero-point CM kinetic energies
have already been proposed.\cite{AMD,FMD,Kiderlen}\ 
As for the calculation of the energy dispersion, 
we remove these effects by subtracting the local average velocities and forces
acting on the wave packets,
\beqar
\sigma^2_{E}\ &=&\ \sum_{ij}
                    \bar{\bold{F}'}_i \cdot C_{ij} \cdot \bold{F}'_j\ ,\\
\bold{F}'_i\ &=&\ \bold{F}_i
                   -{\gamma\over n_i}\,\sum_k f_{ik} \bold{F}_k\ ,
\eeqar
where $f_{ij}$ denotes the ``friendship'' function which is used
to subtract the fragment zero-point CM kinetic energies
as described later in Subsec.~\ref{subsec:effec},
$n_i$ is the mass number of the fragment to which the $i$-th nucleon belongs, 
and here we use $\gamma=1$. 
By this subtraction,
the energy fluctuations are correspondingly reduced,
thereby ensuring that the fluctuations disappear
when all the fragments have become cold.

%
%
The second problem concerns the expression of the fluctuation
matrix $\bold{g}$.
If we were to adopt a simple constant matrix $\bold{g}$ 
having only diagonal terms as in the Lipkin model, 
both of 
the translational motion 
and the intrinsic motion of isolated fragments close to their ground states
would be affected by the fluctuation.
The first point can be avoided by adopting a fluctuation matrix of the form,
\beq
\label{eq:g-rough}
g_{ij} \sim \delta_{ij} - 1/A_F \ ,
\eeq
where $A_F$ is the mass number of the fragment to which the nucleon
belongs.
On the other hand, the intrinsic (or relative) motion between nucleons
are not affected when the same fluctuations are added.
Then, we can avoid the second problem by requiring that
all the matrix elements of $\bold{g}$ are similar within an isolated
fragment near the ground state.
In order to satisfy both requirements approximately, 
we employ the following matrix $\bold{g}$ that contains off-diagonal parts
reflecting single-particle overlaps,
\beqar
\label{eq:gmatform}
g_{ij} &=& g_0 
                    \left( \frac{\sigma_{E}}{\hbar\sqrt{A}} \right)^{1/2}
                    \left(
                    \frac{\tilde{f}_{ij}}{\sqrt{q_i q_j}} 
                    -
                    \frac{f_{ij}}{\sqrt{n_i n_j}} 
                    \right)\ ,\\
\tilde{f}_{ij} &=&
\exp\left(
      - \nu_N \left|
		\frac{\w_i}{\sqrt{\nu_i}} - \frac{\w_j}{\sqrt{\nu_j}}
		\right|^2 \right)
    \ ,
\eeqar
where $q_i=\sum_k \tilde{f}_{ik}$.
In this form,
the diagonal part in Eq.~(\ref{eq:g-rough}) is replaced by a narrow Gaussian,
and the constant part for each fragment is replaced by a normalized
friendship function having a larger width.
With this form,  
the matrix elements of $\bold{g}$ become almost zero
and the energy is unaffected by the fluctuation
for fragments close to their ground states.
Here we have introduced one free parameter $g_0$, which governs the overall
strength of the fluctuation matrix.
We determine this free parameter later in Subsec.~\ref{subsec:p-induced}.
%
%
By using this fluctuation matrix, 
the Quantal Langevin force is implemented into AMD as follows,
\beq
\dot{\z}_i =      \frac{i}{\hbar}\bold{F}_i
               +  \beta_{\Hml}
                  \sum_{kl}
                  g_{ik}g_{kl}\bold{F}'_l
               +  \sum_{k}
                  g_{ik}\bold{\zeta}_k\ .
\eeq

%
%
In addition to introducing a stochastic force in the equation of motion,
   the Quantal Langevin model 
   requires that the intrinsic distortion
   of wave packets be taken into account in order to project 
   the state to the appropriate energy shell before making
   any observation.\cite{OR97a,OR97b,OR97c}\ 
In the case where the energy is specified,
the distortion operator $\sqrt{\delta(E-\hat H)}$ is very complicated.
Therefore, we have employed the canonical distortion here.
   This can be done by using the following cooling equation
   to perform an imaginary-time evolution,
\beqar
\label{eq:cool}
\dot{\z}_i &=&  \bold{F}'_i\ ,
\eeqar
until the total energy expectation value reaches the given energy.
Although this cooling process hardly influences the fragment configurations
at the final stage of the reaction, 
it reduces the excitation energy of fragments significantly
and the energy conservation is completely satisfied after this procedure.

\subsection{Effective Interactions} 
\label{subsec:effec}

%
%
We use the following effective interactions in this work.
For the $NN$ interaction,
the effective interaction Volkov No.1~\cite{Volkov}
   with Majorana parameter $m=0.575$ is used.
%
%
The zero-point kinetic energy ($\Kin^{CM}$) is subtracted from the 
Hamiltonian in the same way as in Ref.~\citen{AMD}.
As a result, the total Hamiltonian is modified to the following form,
\vspace*{-0.5cm}
\beqar
\Hml      &=& \bra{\Z} \hat{H} \ket{\Z} -\Kin^{CM} \ ,\\
\Kin^{CM} &=& T_0 A - a_p T_0 ( A - N_F )          \ ,\\
               N_F &=& \sum_i \frac{ 1 }{ n_i m_i }\ ,
\quad
n_i  =  \sum_j f_{ij}\ ,
\quad
m_i  =  \sum_j \frac{ f_{ij} }{ n_j } \ ,\\
f_{ij}    &=& \exp \left( -\nu_t\,\left|
		\frac{\z_i}{\sqrt{\nu_i}} - \frac{\z_j}{\sqrt{\nu_j}}
		\right|^2 \right)\ .
\eeqar 
Here, $T_0 = 3 \hbar^2 \nu/2M$ is the zero-point kinetic energy of a fragment,
$A$ is the total mass number, and $N_F$ is the number of fragments 
appearing in the dynamical simulation.
The parameters involved in the $\Kin^{CM}$ are determined as 
$\nu_t$ = 0.1 and $a_p$ = 0.772 in order to reproduce the 
experimental data of binding energies of normal nuclei as shown
in Table~\ref{table:be}.

\begin{table}
\caption{Binding energies of single hyper nuclei}\label{table:be}
\begin{center}
\begin{tabular}{cccccccc}
\hline   
\hline
\multicolumn{2}{l}{Core nucleus}
&\nuc{3}{H}  &\nuc{4}{He} &\nuc{7}{Li} &\nuc{8}{Be} &\nuc{11}{B}&\nuc{12}{C}\\
\hline
$B(\hbox{Core})$ & cal.(MeV)
&  10.63     & 28.30       & 39.66     & 56.73      & 74.51     & 92.35	\\
& exp.(MeV)
&  ~8.48     & 28.30       & 39.25     &  56.50     & 76.21     & 92.17	\\
\hline 
\multicolumn{2}{l}{Hypernucleus}
&\nucl{4}{H}&\nucl{5}{He}&\nucl{8}{Li}&\nucl{9}{Be}&\nucL{12}{B} & \\
\hline
$B(^A_\Lambda Z)$ & cal.(MeV)
&  12.41     & 31.43      &  45.32     & 60.99      & 87.15     &       \\
 &exp.(MeV)
&  10.52     & 31.42      &  46.05     & 63.27      & 87.58     &       \\
\hline
$S_\Lambda$& cal.(MeV)
&  1.78      & 3.13        &  5.66     & 4.26       & 12.64     &	\\
&exp.(MeV)
&  2.04      & 3.12        &  6.80     & 6.77       & 11.37     &	\\
\hline
\multicolumn{2}{l}{Fragmentation to}
&            &             &\nuc{3}{H}+\nucl{5}{He}
                                       &\nuc{4}{He}+\nucl{5}{He}&& \\
$Q$ values
&cal.(MeV)&  &             & 3.26      & 1.26       &           &	\\
&exp.(MeV)&  &             & 6.15      & 3.55       &           &	\\
\hline   
\end{tabular}
\end{center}
\end{table}

%
%
Other interactions ($N \Lambda$, $\Lambda \Lambda$) are assumed 
   to be attractive Gaussian potentials, 
\beqar
        v_{\Lambda \Lambda}
        &=& -90.12\exp(-0.935r^2)\ , \\
        v_{\Lambda N}
        &=& -43.62(0.1 - 1.0P_{\sigma} + 0.5P_r)\exp(-0.935r^2)\ .
\eeqar
The range of the $N\Lambda$ interaction is the same as that of 
   two-pion exchange\cite{Lam-rang} \ 
and other parameters are chosen to fit the 
   experimentally known hypernuclear binding energies of 
   \nucl{4}{H}, \nucl{5}{He}, \nucl{8}{Li}, \nucl{9}{Be}, and \nucL{12}{B}
   within AMD wave functions as shown in Table~\ref{table:be}.
The largest ambiguity lies in the $\Lambda\Lambda$ interaction,
since the information on double hypernuclei is very scarce.
We choose the parameter of the $\Lambda\Lambda$ interaction to fit
the $\Delta B_{\Lambda \Lambda}$ of \nucll{13}{B}(=4.9 $\pm$ 0.7 MeV) 
obtained in KEK E176 experiment.\cite{DoubleC,Stopped}\ 

In Table~\ref{table:be}
we also show the $\Lambda$ separation energies 
and the $Q$ values of some fragmentations that are relevant to this work.
These energies are very important,
since they largely affect the probabilities of $\Lambda$ emission
and decay to fragments in the dynamical evolution.
The adopted effective interactions reproduce the
observed separation energies within about 2 MeV,
although 
the separation energies of light clusterized nuclei
(\nucl{8}{Li} and \nucl{9}{Be}) and their $Q$ values of fragmentations 
(\nucl{8}{Li} $\to$ \nuc{3}{H}   + \nucl{5}{He},
 \nucl{9}{Be} $\to$ \nuc{4}{He}  + \nucl{5}{He})
are underestimated a little more
because of the limitation of the AMD wave function.
We will discuss the consequences 
of this underestimate later.
%

\subsection{Two-Body Collisions}
\label{subsec:coll}

%
%
Two-body collisions are also included by using the physical coordinate
   $\w$.\cite{AMD}\ 
The following elementary collisions are included in our calculations,
\beqar
(1)\ &~&\ N+N \to N+N\ ,\\
(2)\ &~&\ N+\Lambda \to N+\Lambda \ , \\
(3)\ &~&\ \Lambda+\Lambda \to \Lambda+\Lambda\ ,
\eeqar
where $N=n,p$.
The $NN$ cross sections are taken from the experimental data.
The cross sections involving $\Lambda$ are calculated
by using the Nijmegen model D.\cite{NijmegenDYN} \ 
The Nijmegen model D gives the coupling constants for $YN$ interactions
and the cross sections for $N+\Lambda \to N+\Lambda$,
assuming a short-range core ($r_c \sim 0.5$ fm) 
in all the baryon-baryon channels.
The cross sections for $\Lambda+\Lambda \to \Lambda+\Lambda$ 
can be obtained by extending the model D to SU(3) symmetry
and keeping the hard-core radius as $r_c = 0.5$ fm.

\subsection{Determination of the fluctuation matrix $\bold{g}$
through proton-induced reactions}
\label{subsec:p-induced}

\begin{figure}
\begin{minipage}{15cm}
\vspace*{-0.5cm}
\hspace*{-1cm}
\geteps{18cm}{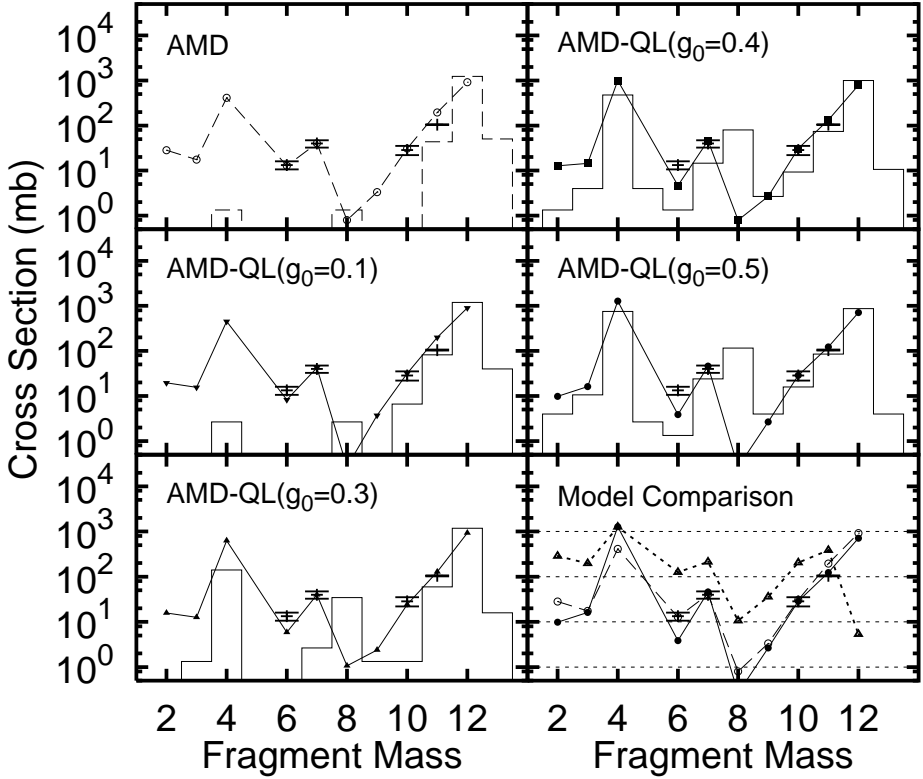}
\end{minipage}
\caption{Calculated fragment mass distribution in the reaction
\nuc{12}{C}+p at 45 MeV
compared with the experimental data.
The histograms show the calculated mass distribution 
at the end of the dynamical stage,
while the lines with points show those after the subsequent statistical decay.
Dashed and solid lines refer to AMD and AMD-QL, respectively,
using $g_0= 0.1 \sim 0.5$.
In the bottom-right panel,
we compare the results of AMD and AMD-QL ($g_0=0.5$)
with the Cascade results starting from pure compound nucleus
formation $\nuc{13}{N}^*$ (dotted line).
The maximum impact parameter is taken to be 6.5 fm,
which corresponds to a cross section of 1300 mb.
The error bars indicate the experimental values.\protect\cite{p+C12}
}
\label{fig:p-induced}
\end{figure}

Until now, we have left one free parameter $g_0$ which gives 
the overall strength of the Quantal Langevin force.
Although this parameter does not affect the statistical properties
at equilibrium, 
it gives the time scale of the relaxation to equilibrium.
Thus it has some effect on the dynamics, 
especially in preequilibrium processes such as fast emission of particles
or fragmentation.
%
In this study,
we determine the allowed range of the strength parameter $g_0$
in a phenomenological fashion,
by analyzing the fragment production cross sections of the proton-induced
reaction on \nuc{12}{C} at $E_p=$ 45 MeV,
which has a similar excitation energy (43.5 MeV) as that of  
$\Xi^{-}$ absorption at rest on \nuc{12}{C} (39.5 MeV).
%
%

The dynamical simulations are performed 
with AMD and AMD-QL up to the time 200 fm/c.
The decay of the excited fragments produced during the dynamical stage 
are then treated by the multi-step binary statistical decay model 
denoted Cascade.\cite{Casc} \ 
In this manner, we obtain the fragment mass distribution.

First we investigate how normal AMD combined with Cascade works 
in describing the proton-induced reaction on \nuc{12}{C} at 45 MeV.
In the upper left panel of Fig~\ref{fig:p-induced},
we show the results of AMD (histogram) and AMD plus Cascade (open circles)
calculation for the fragment mass distribution of \nuc{12}{C}+p at 45 MeV. 
At first glance it seems that the experimental data~\cite{p+C12}
is reasonably reproduced.
However, in the preequilibrium dynamical stage, which is described by AMD,
the fragmentation hardly occurs. 
One nucleon emission from \nuc{12}{C},
inelastic excitation of \nuc{12}{C},
and formation of the compound nucleus \nuc{13}{N}$^{*}$
are dominant, 
and most of the intermediate-mass fragments
are formed at the statistical decay stage,
as already discussed by Takemoto et al.~\cite{AMD2}\ 
This description may not be reasonable
by the following two reasons.
First, the survival probability of the compound nucleus \nuc{13}{N}$^*$
at the end of dynamical simulation seems too large.
The life time of the compound nucleus \nuc{13}{N}$^*$
at $E^* = 43.5$ MeV is around $\tau \sim$ 65 fm/c in Cascade,
and the compound-nucleus formation cross section is at most 300 mb,
assuming that the grazing impact parameter of around 3 fm.
Therefore, at the time $t=$ 200 fm/c when the dynamical calculation ends,
around 5\% of the compound nuclei \nuc{13}{N}$^*$ can survive,
which corresponds to 15 mb by using the above estimate
and well below the calculated value with AMD ($\sim 50$ mb).
The second is the lack of the preequilibrium emissions.
Basically Cascade describes low-energy statistical
decay of equilibrated excited nuclei
and the angular distribution of the ejectiles
is isotropic, or at least forward-backward symmetric.
However, at these incident energies
the fragment angular distribution shows strong
forward-backward anisotropy,
as shown in the \nuc{9}{Be}/\nuc{11}{B} + p  experimental data,\cite{p+B11}\ 
caused by direct-like fragment production.

Next we analyze \nuc{12}{C} + p at 45 MeV with AMD-QL plus Cascade 
and seek to determine the allowed range of the parameter $g_0$
by comparing the calculated fragment mass distribution with data.
In Fig.~\ref{fig:p-induced},
we show the results of the AMD-QL (plus Cascade) calculation for 
the fragment mass distribution of \nuc{12}{C} + p at 45 MeV
with $g_0$=0.1 $\sim$ 0.5.
After the statistical decays
we cannot see significant differences among the results
of the AMD and AMD-QL calculations, and all of them well reproduce the data.
However, the mass distribution at the dynamical stage
(histograms in Fig.~\ref{fig:p-induced})
changes drastically when the value of $g_0$ is changed.
As the fluctuation strength is increased from $g_0$=0 (AMD) to $g_0=0.5$,
the compound nuclear survival probability decreases
and dynamical fragmentation (mainly to $\alpha+$\nuc{8}{Be} and 3$\alpha$)
occurs more frequently.
We should note that AMD-QL can describe fragmentation
to intermediate-mass fragments at low excitation
before statistical decay by the effects of fluctuations
around classical paths given by Quantal Langevin force.
So, the mass distributions before and after Cascade 
become closer, 
and direct production of ground-state fragments is 
partly described with AMD-QL.
On the basis of this analysis of proton-induced reactions,
we find the range $g_0 \leq 0.5$ for the fluctuation strength parameter
is suitable for reproducing the fragment mass distribution after Cascade.
In addition, from the consideration of the life-time of compound
nuclei and the necessity of preequilibrium fragment emission, 
the values around $g_0 \sim 0.5$ seems preferable.

It is also possible to describe the reaction \nuc{12}{C} + p at 55 MeV
with the same value of the strength parameter $g_0$.
Detailed analysis of proton-induced and heavy-ion reactions with AMD-QL 
will be reported elsewhere.

\newpage

\section{Results for the $\Xi^{-}$ absorption reaction}\label{sec:results}

We now apply the AMD-QL simulation, and the subsequent Cascade process,
to the absorption of $\Xi^{-}$ at rest on the nucleus \nuc{12}{C}.

\subsection{Initial Condition}
\label{subsec:results:initial}

%
%
%
%
The initial wave function of the $\Xi^-$ hyperon is calculated
   by assuming the interaction between $\Xi^{-}$ and \nuc{12}{C} to be
   the following Woods-Saxon potential and the Coulomb potential,
\beqar
U(r) &=& V_0 \left[ 1+ \exp({r-R \over a}) \right]^{-1} + U_{Coul}(r)\ ,
\eeqar
where $R = r_0 A^{1/3}$ denotes the nuclear size, and 
   $V_0$ is the potential depth.
Here, we ignore the effects of the imaginary part of $V_0$
on the wave function.
The size and the diffuseness parameters are taken as
$r_0 = 1.14$ fm and $a=0.65$ fm, respectively.

%
%
Once the $\Xi^-$ wave function is known,
   the absorption point of $\Xi^-$ is calculated by the density overlap 
   between $\Xi^-$ and the protons in \nuc{12}{C},
\beq
\label{Overlap}
{\del w \over \del t}
        \propto \rho_p(\r) \rho_{\Xi^-}(\r)\ .
\eeq
It is noteworthy that this density overlap does not depend on the 
   radial quantum number $n$ for higher $\ell$ if it is normalized,
   since it is sensitive only to the behavior at the nuclear surface. 
We assume that $\Xi^-$ always reacts with a proton in \nuc{12}{C} 
   and two $\Lambda$ hyperons are produced at the points where the $\Xi^-$
   and proton existed.
We neglect the elementary reactions involving the $\Sigma$ hyperon,
since those channles are closed.
This elementary reaction is described by ordinary collision term based on
   the canonical variable $\bold{w}$.\cite{AMD} 
The spins of the produced $\Lambda$ hyperons are determined statistically.
%
%
The energy of the $\Xi^{-}$ absorption on \nuc{12}{C} in the experiment
   in which twin hypernuclei was discovered~\cite{TwinA,TwinB} was 
   well reproduced by a shallow potential ($V_0=-16$ MeV)
   and two events of twin hypernuclear formation,
   \nuc{12}{C} + $\Xi^{-}$ $\rightarrow$ \nucl{9}{Be} + \nucl{4}{H},
   are interpreted as the reactions in which $\Xi^{-}$ is absorbed
   from a $2p$ state.\cite{TwinB,Tadokoro} \ 
In this study, we assume that $\Xi^{-}$ is absorbed from a
(1 or 2)$p$ state
   and $V_0=-16$ MeV. 

\subsection{Hyperfragment Mass Distribution}\label{subsec:results:massdist}

We have made simulation calculations by using AMD and AMD-QL
($g_0$=0.1, 0.3, 0.4, and 0.5)
up to $t= 200$ fm/c.
A few thousand events were generated
for each value of the strength parameter $g_0$.
In the case of AMD-QL, we have taken the 
effects of intrinsic distortion of wave packets into account
at the end of the reaction by the imaginary-time evolution
given in Eq.~(\ref{eq:cool}). 
In this procedure, we have ignored those events 
in which the total energy expectation value has not reached the
specified energy $E_0$.
%
%
In the statistical decay stage, 
the fragment mass and isotope distributions are determined essentially by
the $Q$ values of the decay, the Coulomb barrier,
and the spin-degeneracy factor.
Therefore,
we have used the experimentally known hypernuclear binding energies,
if available.
If not, the separation energies calculated
by using the AMD wave functions are adopted.
On the other hand, the fragment excitation energies used as the inputs 
for the statistical decay are calculated by subtracting the calculated 
ground-state energies
in order to avoid the spurious effects arising
from the difference between the calculated binding energies in the model
space and the experimental values.

\begin{figure}[htb]
\begin{minipage}{15cm}
\hspace*{-2.5cm}\geteps{20cm}{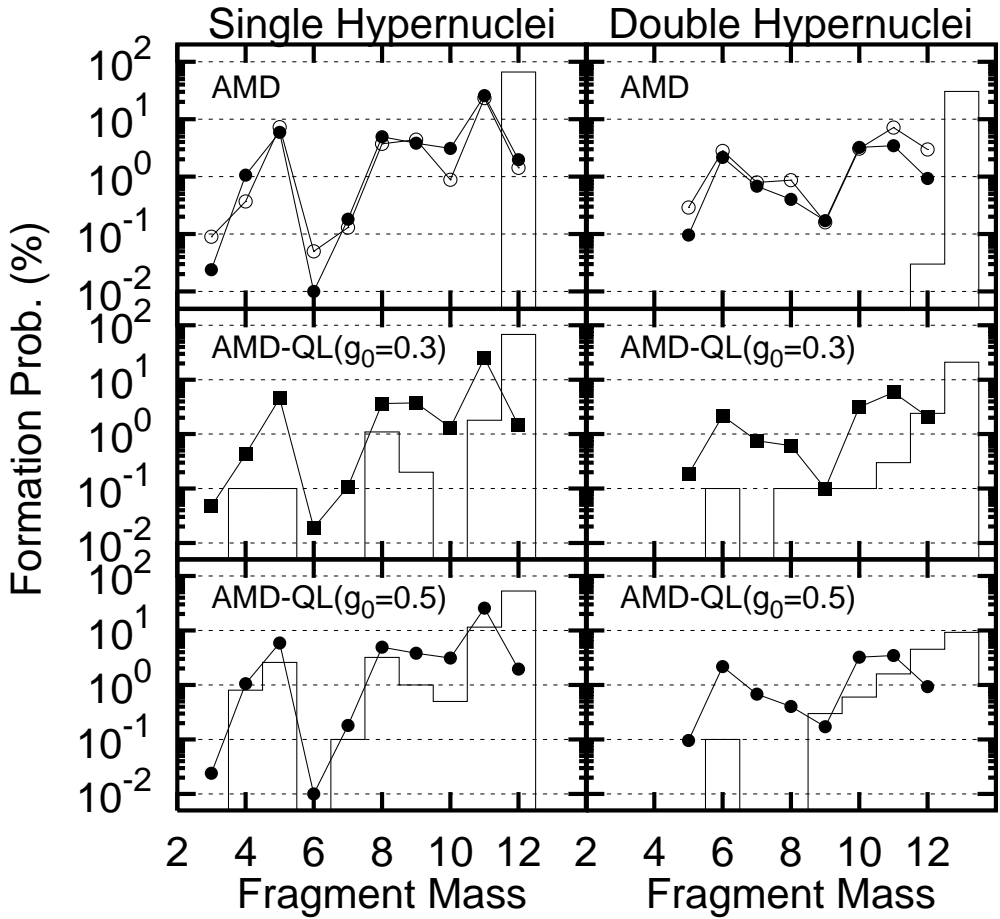}
\end{minipage}
\caption{
Calculated single and double hypernuclear mass distributions 
in $\Xi^-$ absorption reactions at rest on \nuc{12}{C}.
Histograms in the upper, middle, and lower panels
show calculated results
with AMD, AMD-QL ($g_0=0.3$), and AMD-QL ($g_0=0.5$)
after the dynamical stage,
respectively.
Open circles (upper panel), squares (middle),
and solid circles (lower)
show the results with
AMD, AMD-QL ($g_0=0.3$), and AMD-QL ($g_0=0.5$)
followed by Cascade,
respectively.
For comparison,
the upper panel also includes the results for AMD-QL ($g_0=0.5$)
followed by Cascade.}
\label{fig:ximass}
\end{figure}

In Fig.~\ref{fig:ximass},
we show the calculated single and double hyperfragment mass distribution
in $\Xi^-$ absorption reaction at rest on \nuc{12}{C}
with AMD and AMD-QL combined with Cascade.

%
%
For the single-hyperfragment mass distribution,
the situation is similar to the case of the \nuc{12}{C}+p reaction:
With AMD, we only obtain formation of the \nucL{12}{B}$^*$ compound
nucleus by one $\Lambda$ evaporation in the dynamical stage.
Almost all the single-hyperfragments are formed through
the statistical decay of $\nucL{12}{B}^*$ and $\nucll{13}{B}^*$.
%
On the other hand, with AMD-QL, fragmentations occur more frequently,
mainly to \nucL{11}{B}, \nucL{11}{Be}, \nucl{8}{Li}, and \nucl{5}{He}, 
which have relatively large binding energies
and have stable fragmentation partners after one $\Lambda$ emission.
As the fluctuation strength $g_0$ is increased, dynamical fragmentation
enhances.
In addition, we can see some signals of the dynamical effect in the 
single-hyperfragment mass distribution in AMD-QL followed by Cascade.
For instance, $A$=10 ($\nucL{10}{Be}$ and $\nucL{10}{B}$) fragments 
are created mainly through one nucleon (or $\Lambda$) evaporation from
dynamically produced $A$=11 single (or double) hyperfragments in
AMD-QL followed by Cascade,
but these processes are not possible in AMD plus Cascade.

%
%
For the double hyperfragment mass distribution, the effects of quantal
fluctuations of AMD-QL are more evident.
For instance, the production rates of $A$=11 and 12 double hyperfragments
calculated with AMD-QL followed by Cascade 
are much smaller than those 
of AMD plus Cascade.
This difference comes from the compound nucleus $\nucll{13}{B}^*$
formation probability at the end of the dynamical stage.
(Although the above difference looks small on a logarithmic scale,
 it is significant on a linear scale,
 see Fig.~\ref{fig:doubletwin}.)
This formation probability is about 30\% in AMD,
while it decreases to around 10\% for $g_0$= 0.5 in AMD-QL.
In other words, the $\Lambda$ emission probability is enhanced in AMD-QL.

The fact that the formation of long-lived double-hyperon compound nuclei
occurs so frequently can be understood as follows. 
In the initial stage of the $\Xi^{-}$ absorption reaction,
   a proton in \nuc{12}{C} 
is converted to a $\Lambda$ and the residual nucleus becomes \nuc{11}{B}$^*$.
The separation energy of this proton is around 16 MeV,
and \nuc{11}{B}$^*$ is formed with about 6 MeV excitation in AMD.
As a result, a large part of the released energy in the elementary 
   reaction of $\Xi^- p \to \Lambda \Lambda$ (28.3 MeV) is expended
and the energy available to the 2$\Lambda$ system is very small 
($\sim$ 3 MeV/particle).
In AMD, the quantal fluctuations associated with wave packets are neglected
and the motion of $\Lambda$ is limited by the given small energy.
Therefore, the formation of long-lived double-hyperon compound nuclei
occurs more frequently in AMD calculation.

\begin{figure}[htb]
\vspace*{-3cm}
\centerline{\geteps{6cm}{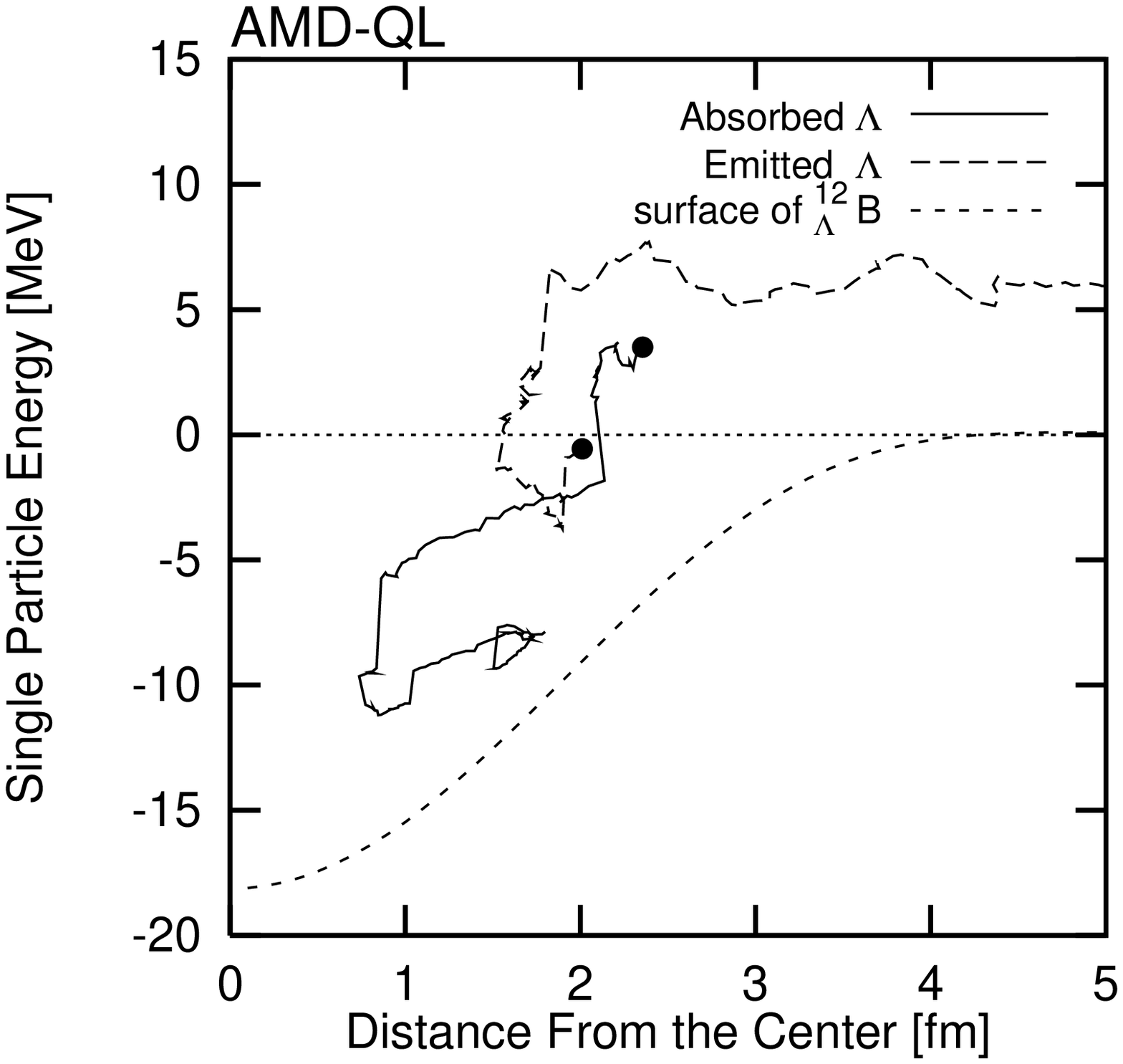}\hspace*{5mm}
\geteps{6cm}{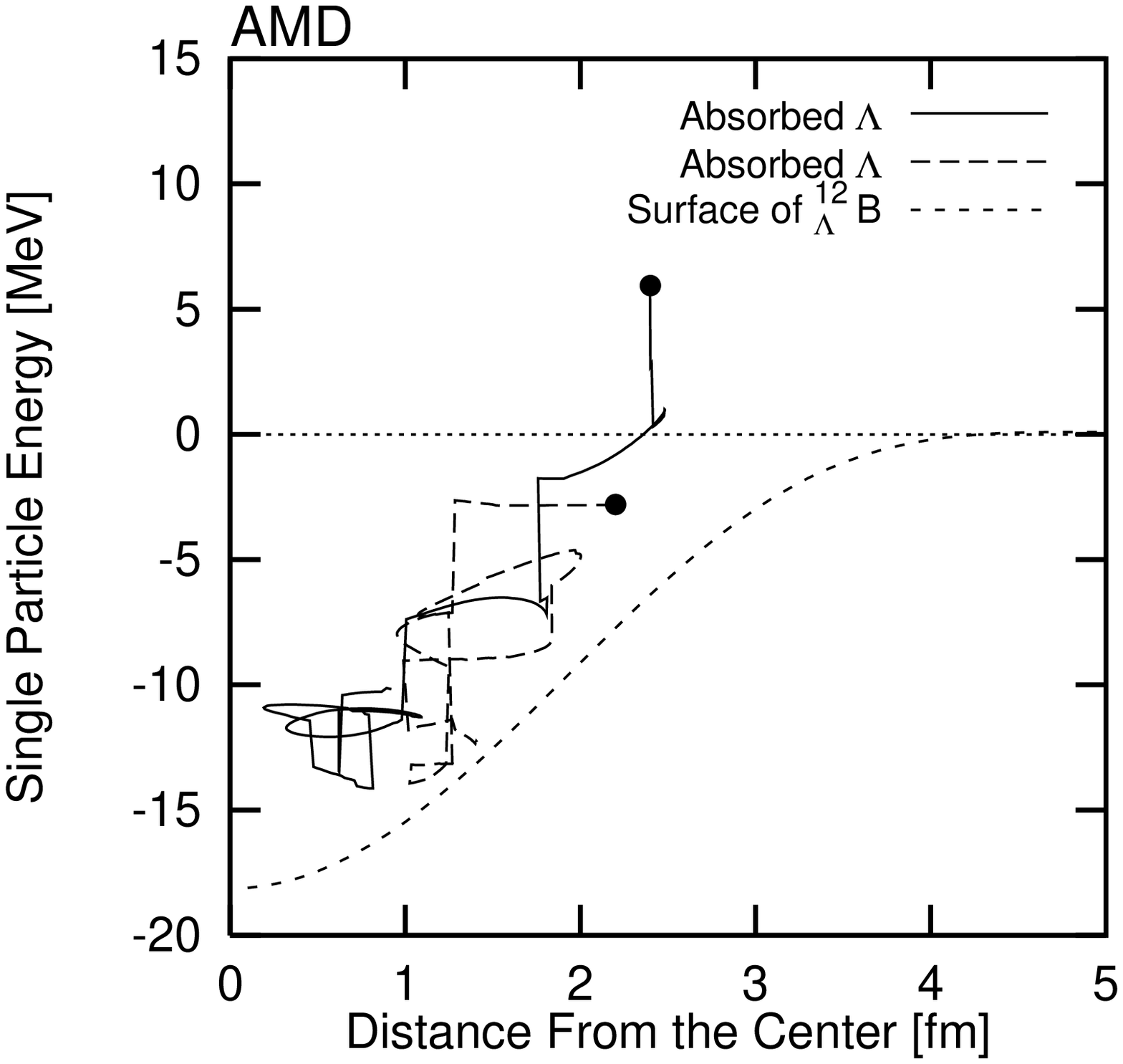}}
\caption{
Typical time development of the single-particle energies
of the two $\Lambda$ hyperons,
as obtained with AMD-QL (left) and AMD (right).
The abscissa is the distance from the center-of-mass of the system 
and the ordinate is the single-particle energy.  
The solid circles indicate the initial position of the hyperons
at the time of the $\Xi^{-}p$ conversion to $\Lambda\Lambda$. 
The single-particle potential energy
of $\Lambda$ with respect to \nucl{12}{B}	
is shown by the dotted curve.		
In the AMD-QL simulation, 
the hyperons are kicked by the random force
and one is emitted (dashed trajectory in the left panel).
On the other hand, in AMD simulation,
even the hyperon that initially had a positive single-particle energy
is absorbed into the hyperon compound nucleus
due to the energy lost by the collision with a resident nucleon
(solid trajectory in the right panel).
}
\label{fig:emission}
\end{figure}

%
%
In order to show the mechanism of $\Lambda$ evaporation in the
dynamical stage of AMD-QL, we analyze the dynamical evolution of the 
single-particle energy of each of the two produced $\Lambda$ hyperons.
Figure~\ref{fig:emission} displays a typical time development 
of two $\Lambda$ hyperons in a projection showing the single-particle energy
and the distance from the center of the nucleus.
The left (right) panel is for AMD-QL (AMD) and
the initial projections are shown by solid dots.
Since 
 in AMD-QL the Quantal Langevin force continually kicks the particles,
one of the hyperons is occasionally emitted towards the continuum region,
leading to its evaporation from
the highly excited \nucll{13}{B}$^*$ compound nucleus.
In AMD-QL, this type of $\Lambda$ evaporation process dominates
and the double hyperfragment formation is suppressed. 
By contrast,
with the AMD simulation the hyperons are steadily losing their energy,
leading to their eventual absorption into 
the compound nucleus. 
In this case two hyperons lose most of their single-particle energy
as a result of collisions with the nucleons.
In AMD this type of energy loss occurs frequently 
and double-hyperon compound nucleus formation therefore becomes more frequent.

It is interesting to compare the \nuc{12}{C} + $\Xi^{-}$ reaction,
in which two $\Lambda$ hyperons are produced, 
with the \nuc{12}{C+p} reaction in which all particles are nucleons.
In the latter case
we cannot tell from the mass distribution alone
whether the particle was emitted in the dynamical stage 
or in the statistical stage
because
the incident (leading) particle and the residents of the target
nucleus are identical.
Therefore, the mass distribution does not depend on whether
the ejectile is the leading particle or arises from the target.

On the other hand, in the $\Xi^-$ absorption reaction,
the leading two hyperons are different from nucleons
and we can tell whether the ejectile is one of those or not.
During the dynamical (or preequilibrium) stage,
the probability for leading particle $\Lambda$ to escape
is larger than that for the resident nucleons.
However, there is no difference once equilibrium has been achieved.
Therefore, we may be able to clarify the contribution
of the dynamical fragmentation and statistical one 
from the $\Lambda$ emission probability of the 
$\Xi^-$ absorption reaction.

\subsection{Event Type Analysis}\label{subsec:results:type}

%
%
Table~\ref{table:g-dep} shows the dependence
on the fluctuation strength parameter $g_0$
of the relative abundance of different event types;
double-, single-, and twin-hyperfragment formation,
as well as events with no hyperfragment formation.
The AMD results correspond to $g_0$=0. 
As described in the previous subsection, 
quantum fluctuations enhance preequilibrium emission of
one $\Lambda$ particle,
while suppressing formation of the double-hyperon compound nucleus
$\nucll{13}{B}^*$ (second column in Table~\ref{table:g-dep}).
This feature also applies to emission of other particles.
In the dynamical stage,
the double hyperfragment formation probability monotonically decreases
and two-$\Lambda$ emission events (no hyperfragment formation)
grows, as $g_0$ increases.
For single hyperfragment formation,
it first grows following enhanced one $\Lambda$ emission,
but it saturates and decreases again when two $\Lambda$ emission
becomes visible.

\begin{table}[htbp]
\caption{
Dependence on the fluctuation strength $g_0$
of the event type abundance (\%)
for double, single, twin, and no hyperfragment formation.
Pure statistical decay results with Cascade and the statistical
multifragmentation model~\protect\cite{Compound}\ are also shown
for comparison. In the results of the statistical multifragmentation model,
the sum of single and no hyperfragment formation rates is shown
with superscript $a$.
}
\label{table:g-dep}
\begin{center}
\begin{tabular}{crrrrrrrrr}
\hline
\hline
     &\multicolumn{5}{l}{Dynamical(\%)} & \multicolumn{4}{l}{+Cascade(\%)}\\
$g_0$
    & $\nucll{13}{B}^*$
           & Double & Single & Twin & No Hyp. 
                                        & Double & Single & Twin & No Hyp.\\
\hline
0.0
(AMD)&30.3  &30.3   &66.3   &0.0  & 3.4 &18.1   &39.7   &0.9  &41.3\\
  0.1&29.6  &30.1   &66.1   &0.0  & 3.8 &18.1   &38.4   &0.8  &42.7\\
  0.3&21.0  &24.1   &71.3   &0.1  & 4.5 &15.1   &38.0   &0.7  &46.1\\
  0.4&16.4  &20.4   &72.0   &0.3  & 7.3 &12.9   &42.2   &0.9  &44.0\\
  0.5& 9.2  &16.3   &70.9   &0.9  &11.9 &11.4   &44.0   &1.2  &43.4\\
\hline
Cascade
     &100.0 &100.0  & ---   & --- & --- &58.8   &24.5   &3.1  &13.6\\
Compound\protect\cite{Compound}
     &100.0 &100.0  & ---   & --- & --- &66~~   &20$^a$ &14   &---$^a$\\
\hline
\end{tabular}
\end{center}
\end{table}

As the fluctuation strength $g_0$ is increased,
the probabilities after the statistical decay stage
generally become closer to those after the dynamical stage.
This feature is understood as arising from the quantum statistical nature:
In AMD-QL with a sufficient fluctuation strength,
a large part of dynamically produced light fragments 
have excitation energies that are sufficiently small
to ensure survival during the statistical decay stage.
This is because the lower energy components in the wave packet are
emphasized by quantum statistics,
as realized in the calculation 
of the intrinsic distortion of wave packets.
Thus the difference between total production rates of hypernuclei
at the end of the AMD-QL simulation
and after the subsequent statistical decay becomes smaller,
while the Cascade after-burner is still important in describing
the decay of heavier fragments, such as 
\nucll{13}{B}, \nucll{12}{B}, \nucll{12}{Be} and $\nucL{12}{B}$.

%
The combined results of AMD and AMD-QL with Cascade
show that all of these probabilities are in the allowed range between
the lower and upper limits experimentally estimated
at 90\% confidence level, shown in Table~\ref{table:ULLimit},
and the single hyperfragment formation probability is
in the roughly estimated range (26-73\%).
These things apply at any $g_0$ value.
In addition, with a larger fluctuation strength,
the double (twin) hyperfragment formation probability decreases (increases)
and becomes closer to the rough estimate 3-9\% (6-18\%).

\begin{figure}[htbp]
\begin{minipage}{15cm}
~\hspace*{-0.5cm}\geteps{8cm}{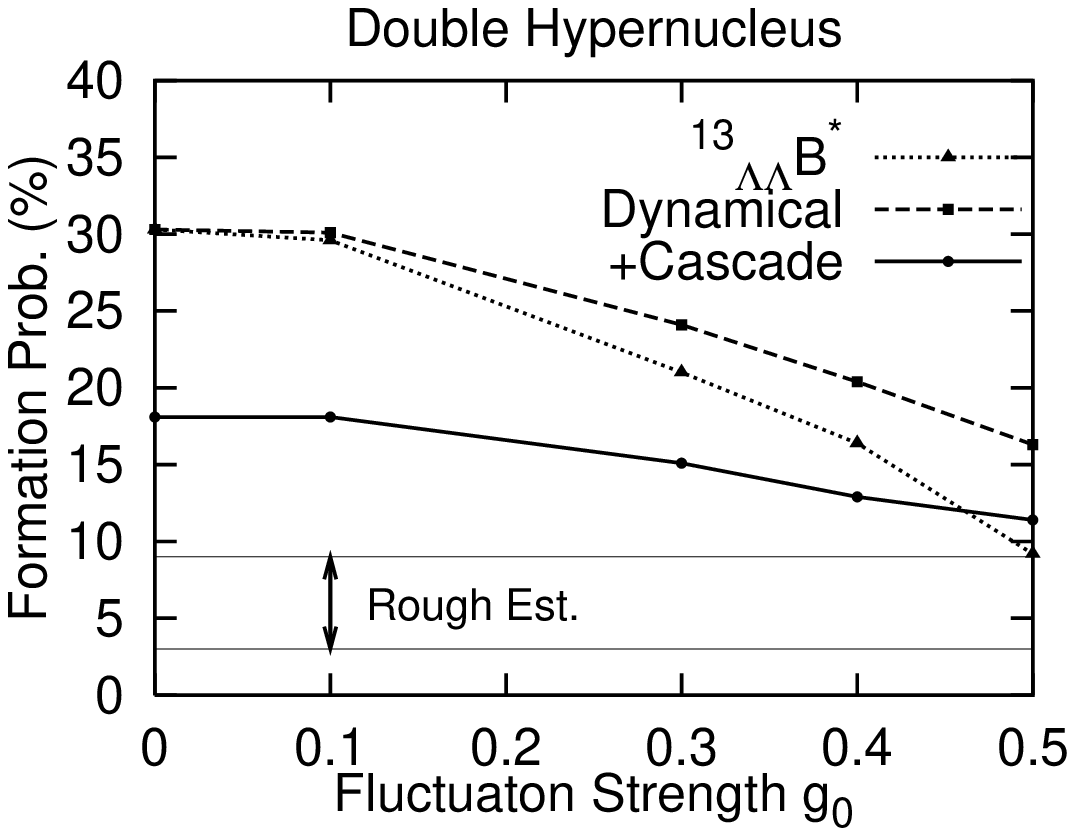}\hspace*{-1cm}\geteps{8cm}{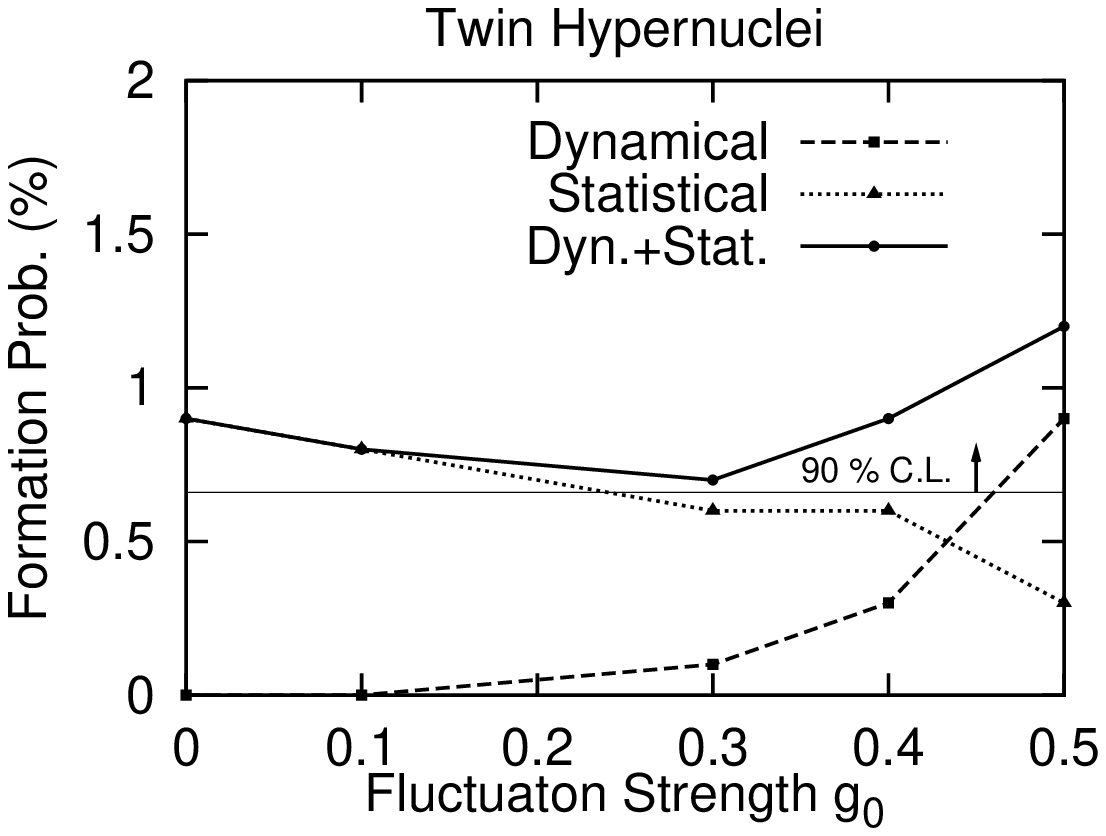}
\end{minipage}
\caption{%
Double- and twin-hypernuclear formation probabilities
as functions of the fluctuation strength parameter $g_0$.
Dashed and solid lines connect the results
before and after statistical decays, respectively.
The dotted curve in the left panel shows the probability of forming
the double-hyperon compound nucleus ($\nucll{13}{B}^*$),
and the dotted curve in the right panel denotes the probability
of twin hyperfragment formation purely from the statistical decay
of the excited double hypernucleus.
The experimental estimates are shown by thin solid lines.
}
\label{fig:doubletwin}
\end{figure}

%
%
%
In Fig.~\ref{fig:doubletwin},
we show the calculated double- and twin-hypernuclear
formation probabilities as a function of the fluctuation strength $g_0$.
At small values of $g_0$, double hypernuclei are calculated to
be formed with a large probability of around 20\%.
When we incorporate quantum fluctuations, 
this probability decreases to around 10\% at $g_0$=0.5
due to the enhanced emission of $\Lambda$ in the dynamical stage.
On the other hand,
the twin formation probability first becomes smaller at small $g_0$ values
and then grows again for $g_0 > 0.3$.
Within the model calculation made here,
this behavior is governed by the competition between
two different formation mechanism of twin hypernuclei;
the statistical formation from the double-hyperon compound nucleus
and dynamical fragmentation.
At small $g_0$ values, the former dominates twin events
and no dynamical formation of twin hypernuclei can be seen.
The probability of the latter grows
as a linear or quadratic function of $g_0$,
and becomes dominant at large quantum fluctuations,
although the twin hyperfragment formation probability is
still significantly smaller than the rough estimate.

\begin{wrapfigure}{r}{\halftext}
\vspace{-2.5cm}
\centerline{\phantom{MMM}\geteps{5cm}{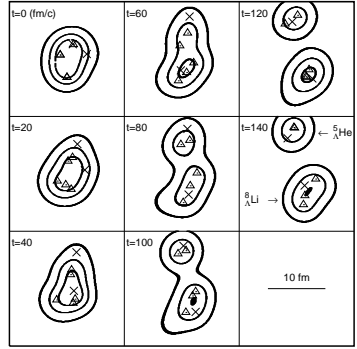}}
\caption{%
Time evolution of the matter density in a twin hyperfragment 
formation event, 
\nuc{12}{C} + $\Xi^{-}$ $\to$ \nucl{8}{Li} + \nucl{5}{He},
calculated with AMD-QL.
Crosses and triangles indicate the 
$\Lambda$ and proton positions, respectively.
Neutrons are omitted for simplicity.}
\label{fig:denstwin}
\end{wrapfigure}

%
%
In Fig.~\ref{fig:denstwin} we show the density evolution of 
   a twin hyperfragment (\nucl{8}{Li}+\nucl{5}{He}) formation 
   event obtained with the AMD-QL dynamical simulation.
In the initial state,
a highly excited \nucll{13}{B}$^*$ compound-like state is formed. 

Around 80 fm/c the compound system starts to fission,
leading to the formation of the hypernuclei
\nucl{8}{Li} and \nucl{5}{He} at very small excitation.
This twin fragmentation occurs over a relatively long time scale
between direct and statistical decay processes.
In addition to twin hyperfragments, frequent light-fragment emission
(such as \nuc{4}{He}, \nuc{3}{H}, and \nuc{7}{Li})
appear in the dynamical stage of AMD-QL.
With quantum fluctuations included we can obtain this variety of
fragmentations during the dynamical stage, before the statistical decay.

\subsection{Exclusive Channels}\label{subsec:results:exclusive}

\begin{table}[htbp]
\caption{
Calculated production rates with AMD and AMD-QL ($g_0=0.5$),
followed by Cascade,
for double, single, and twin hypernuclear formation.
The results of direct reaction theory~\protect\cite{YI97}\ 
are also shown for comparison.
For the value of $\nucLL{5}{H}+2\alpha$ in direct reaction theory
we take the rate of the \nucLL{5}{H}+\nuc{8}{Be}\ channel.
}
\label{table:event}
\begin{center}
\begin{tabular}{llrrrrr}
\hline
\hline
\multicolumn{2}{l}{Channel}
	&\multicolumn{2}{c}{AMD} &\multicolumn{2}{c}{AMD-QL($g_0=0.5$)}
				& Direct~\protect\cite{YI97}\\
& & Dynamical($\%$) & +Cascade($\%$) & Dynamical($\%$) & +Cascade($\%$) & \\
\hline
\multicolumn{2}{l}{Double Total}    &30.3 &18.07 &16.3  &11.39  & \\ 
&\nucll{13}{B}                      &30.3 & ---  & 9.2  & ---   & \\ 
&\nucll{12}{B}+n                    & --- & 2.00 & 2.4  & 0.61  & 3.96 \\ 
&\nucll{12}{Be}+p                   & 0.03& 0.96 & 2.1  & 0.34  & 2.43 \\ 
&\nucll{11}{B}+2n                   & --- & 1.57 & 0.4  & 1.00  & \\ 
&\nucll{11}{Be}+d                   & --- & 0.26 & 0.3  & 0.09  &  0.59 \\ 
&\nucll{11}{Be}+pn                  & --- & 5.27 & 1.0  & 2.88  & \\ 
&\nucll{10}{Be}+t                   & --- & 0.12 & 0.4  & 0.03  & 0.14 \\ 
&\nucll{10}{Be}+(dn,pnn)            & --- & 2.91 & 0.2  & 3.17  & \\ 
&\nucll{\blank9}{Be}+tn             & --- & ---  & 0.1  & 0.10  & \\ 
&\nucLL{\blank9}{Li}+\alp           & --- & 0.16 & 0.2  & 0.05  & 0.05 \\ 
&\nucll{\blank8}{Li}+\alp n         & --- & 0.51 & ---  & 0.19  & \\ 
&\nucll{\blank8}{He}+\alp p         & --- & 0.36 & ---  & 0.24  & \\ 
&\nucll{\blank7}{He}                                                            
+(\nuc{6}{Li},\alp d,\alp pn)       & --- & 0.79 & ---  & 0.49  & \\ 
&\nucll{\blank6}{He}+\nuc{7}{Li}    & --- & 0.23 & ---  & 0.13  & 0.05 \\ 
&\nucll{\blank6}{He}+\nuc{6}{Li}+n  & --- & 0.60 & ---  & 0.37  & \\ 
&\nucll{\blank6}{He}+\alp+(t,dn,pnn)& --- & 1.85 & 0.1  & 1.39  & \\ 
&\nucll{\blank5}{H}+2\alp           & --- & 0.29 & ---  & 0.09  & 0.11$^a$ \\
&Other                              & --- & 0.19 &      & 0.23  & \\ 
\hline
\multicolumn{2}{l}{Single Total}    &66.3 &39.68 &70.9  &43.96  & \\ 
&\nucL{12}{B}+\lam                  &66.3 & 1.42 &53.0  & 3.24  & 2.07\\ 
&\nucL{11}{B}+\lam+n                & --- &17.79 & 7.8  &20.90  & \\ 
&\nucL{11}{Be}+\lam+p               & --- & 5.39 & 3.7  & 5.31  & \\ 
&\nucL{10}{B}+\lam+2n               & --- & 0.01 & ---  & 0.01  & \\ 
&\nucL{10}{Be}+\lam+(d,pn)          & --- & 0.79 & 0.5  & 1.52  & \\ 
&\nucl{9}{Be}+\lam+t                & --- & 4.20 & 0.7  & 2.73  & \\ 
&\nucl{9}{Be}+\lam+dn               & --- & 0.03 & ---  & 0.02  & \\ 
&\nucl{8}{Li}+\lam+\alp             & --- & 3.67 & 2.9  & 4.96  & \\ 
&\nucl{7}{Li}+\lam+\alp n           & --- & ---  & 0.1  & 0.10  & \\ 
&\nucl{5}{He}+\lam+\nuc{7}{Li}      & --- & 6.26 & 1.1  & 3.99  & \\ 
&\nucl{5}{He}+\lam+\alp t           & --- & 0.03 & 0.7  & 0.72  & \\ 
&\nucl{4}{H}+\lam+(2\alp,\nuc{8}{Be})                           
                                    & --- & 0.09 & 0.4  & 0.48  & \\ 
\hline
\multicolumn{2}{l}{Twin Total}      & 0.0 & 0.92 & 0.9  & 1.22  & \\
&\nucL{10}{Be}+\nucl{3}{H}          & --- & 0.08 & ---  & 0.01  & \\ 
&\nucl{9}{Be}+\nucl{4}{H}           & --- & 0.16 & 0.3  & 0.04  & 0.22 \\
&\nucl{9}{Be}+\nucl{3}{H}+n         & --- & 0.01 & ---  & 0.01  & \\ 
&\nucl{8}{Li}+\nucl{5}{He}          & --- & 0.06 & 0.3  & 0.33  & 0.21 \\ 
&\nucl{7}{Li}+\nucl{6}{He}          & --- & 0.05 & ---  & 0.01  & \\ 
&\nucl{7}{Li}+\nucl{5}{He}+n        & --- & 0.08 & ---  & 0.07  & \\ 
&2\nucl{5}{He}+t                    & --- & 0.36 & 0.2  & 0.34  & \\ 
&\nucl{5}{He}+\nucl{4}{H}+\alp      & --- & 0.12 & 0.1  & 0.41  & \\ 
\hline
\end{tabular}
\end{center}
\end{table}

Table~\ref{table:event}
shows the calculated rates for various channels
with AMD and AMD-QL ($g_0=0.5$), both followed by Cascade.
We also show the calculated production rates 
obtained by applying direct reaction theory to the absorption
from the $2p$ atomic state of $\Xi^-$ with $V_0=- 16$ MeV.\cite{YI97}\ 
Although they include the production rates to the particle unbound excited
states, it is still valuable to compare the trend.

As can be seen in Fig.~\ref{fig:ximass}, 
double hypernuclei are mainly produced at large masses ($A$=10, 11, and 12),
because of their larger $2\Lambda$ binding energies.
This feature partly explains why it is difficult to detect
double hypernuclei, since the non-mesic decay rate is expected to be
dominant in these heavier double hypernuclei.
One exception is \nucll{6}{He} (lampha), 
whose production rate reaches around 1.5\%.
While we can see some visible differences between the results of
AMD plus Cascade and AMD-QL plus Cascade,
for the reason described in the previous subsection, 
it is natural that the channels with large $Q$ values are
populated abundantly,
since a large part of these double hypernuclei are produced
through the statistical decay of excited double hypernuclei,
$\nucll{13}{B}^*, \nucll{12}{B}^*, \nucll{12}{Be}^*$.

%
%
As for the single hypernuclei,
those with mass number $A$=11 dominate.
Once a $\Lambda$ is emitted, 
the residual nucleus \nucL{12}{B}, which has the largest probability
in the dynamical stage, will be excited less than 25 MeV,
depending on the energy of the emitted $\Lambda$.
These excitation energies are comparable to
the nucleon and $\Lambda$ separation energies
and therefore a single emission is enough
to bring the residue into the particle stable regime.
The abundant single hypernuclei 
which are produced from \nucL{12}{B},
\nucL{11}{B}, \nucL{11}{Be}, \nucl{9}{Be}, \nucl{8}{Li}, and \nucl{5}{He},
have their stable decay partners with small separation energies 
from \nucL{12}{B},
$S_n$= 12.6 MeV, $S_p$= 14.1 MeV (AMD calc.), $S_{\rm t}$=15.8 MeV,
$S_\alpha$= 13.2 MeV, and $S(\nucl{5}{He})$=16.9 MeV,
respectively.
In AMD-QL, \nucL{10}{Be} has also a relatively large probability of 
1.52\%, which is produced through one-nucleon emission from 
dynamically produced \nucL{11}{B} and \nucL{11}{Be},
and deuteron emission from \nucL{12}{B}.
The dominance of heavier single hypernuclei, which mainly decay
in non-mesic ways, 
explains again why we frequently see those events with large energy
release without pion (8 events out of an average of 31.1).

%
%
In twin hypernuclear formation channels,
the difference between AMD and AMD-QL appears most clearly.
While all the twin hypernuclear formation is described
by the statistical decay of $\nucll{13}{B}^*$ in AMD,
it is mainly described in the dynamical stage in AMD-QL
with a sufficient fluctuation strength.
Especially, in the channels
\nucl{8}{Li}+\nucl{5}{He}, 
2\nucl{5}{He}+t, 
and \nucl{5}{He}+\nucl{4}{H}+$\alpha$,
most of the dynamically produced fragments are bound
and therefore survive the statistical decay stage.
Another visible channel in the dynamical stage is \nucl{9}{Be}+\nucl{4}{H},
which is the experimentally observed one.
However, in the \nucl{9}{Be}+\nucl{4}{H} channel,
dynamically produced \nucl{9}{Be} hyperfragments
are usually in their particle-unstable states
which decay into \nucl{5}{He}+$\alpha$.
One reason for this instability 
is the underestimate of the separation energy 
to the \nucl{5}{He}+$\alpha$ channel
within the present effective interaction and AMD wave function.
Another reason might be in the extended density profile of \nucl{9}{Be}.
The fluctuation matrix with the form of Eq.~\raf{eq:gmatform}
is selected so as not to affect the intrinsic motion of well-isolated compact
fragments.
However, \nucl{9}{Be} has an extended $\alpha\alpha\Lambda$ structure,
and so the intrinsic motion will be affected especially
when fragments are separating.

The total probability of twin hypernuclear formation shown here
is 0.9 -- 1.2\%, depending on the fluctuation strength.
When we limit the initial system to be the double-hyperon compound nucleus
$\nucll{13}{B}^*$, 
the Cascade probability resulting in twin hyperfragments is only around
$P_{\rm Twin}(\nucll{13}{B}^*)$= 3\%.
%
On the other hand, 
this probability amounts to
$P_{\rm Twin}(\nucll{13}{B}^*)=$16\%,
within the hyperon compound nucleus picture,\cite{Compound}\ 
which exceeds the Cascade result by a factor of five.
We have found that it is suppressed to be around
$P_{\rm Twin}(\nucll{13}{B}^*)=$ 10\% and 7\% within
a statistical multifragmentation model
similar to that used in Ref.\ \citen{Compound},
when we take account of excited levels of daughter fragments
with  mass number $A \geq 5$
up to either $E^* = E_{\rm thr}+2 (A-4)$ MeV
or infinity, respectively.
Therefore, a factor of two may be explained by a different treatment 
of the excited fragment states at the statistical decay stage. 
The remaining factor 2-3 may be due to differences in decay scheme.
%
%
%

%

\section{Summary and Discussion}\label{sec:sum}

We have investigated single, twin, and double hypernuclear formation
   from $\Xi^{-}$ absorption at rest on \nuc{12}{C}
   by using Antisymmetrized Molecular Dynamics~\cite{AMD}\ 
augmented by the Quantal Langevin force~\cite{OR95,OR97a,OR97b}\
(AMD-QL)
and followed by the multi-step binary statistical decay model 
(Cascade).~\cite{Casc}

The Quantal Langevin treatment ensures that the dynamical evolution
of the system leads towards quantum statistical equilibrium.
Such dynamics can be described by the time evolution of a distribution function
   that satisfies a Fokker-Planck equation,
   and the equivalent time evolution of the complex wave-packet
   parameter $\{\z_i\}$ is governed by a Langevin-type equation of motion.
This model has been shown to work well in a statistical
context; it describes the statistical equilibrium properties of
simple soluble models and finite nuclei well,\cite{OR95,OR97a,OR97b}\
and it has been used to study equilibrium properties
of argon fluids.\cite{OR97c}
Furthermore,
it has met with some success in describing
the fragment yields in Au+Au collisions
at higher energies (100 -- 400 MeV/A) 
where the number of particles are large and the statistical
equilibrium may be achieved.\cite{OR97b}
Therefore, the model may be expected to work well
also in dynamical contexts
where pre- or non-equilibrium aspects are important.

In order to show the validity of the Quantal Langevin model
in a dynamical context, we first applied it to the damping 
of collective motion within a simple soluble Lipkin model. 
It was shown that the damping process is well described
within the Quantal Langevin model,
including its dependence on the quasi-spin degeneracy $N$.
Although it is difficult to treat the interference between specific
energy eigencomponents that agitate the collective motion again
after several periods,
the early evolution coincides well with that of the exact solution.
%

%
%
On the basis of the above success in the Lipkin model, 
we then incorporated the Quantal Langevin force into
the Antisymmetrized Molecular Dynamics (AMD).
In the resulting model, referred to as AMD-QL,
the usual AMD equation of motion for $\{\z_i\}$
is augmented by the inclusion of the quantum fluctuations,
leading to a Langevin-type equation.
In order to achieve reliability in the practical applications,
we have employed a relatively simple parametrized matrix $\bold{g}$
and determined the range of the overall fluctuation strength parameter,
$g_0$, by analyzing the fragment mass distribution of
\nuc{12}{C}+p reaction at 45 MeV,
which has an excitation similar to that of the reaction under consideration,
$\Xi^-$ absorption at rest on \nuc{12}{C}.
This analysis shows that although the fragment mass distribution of AMD-QL
resembles that of AMD after statistical decays,
it is significantly different in the dynamical stage.
In AMD, inelastic excitation of \nuc{12}{C}, compound nuclear formation,
and one-nucleon emission exhaust almost all the events
and no fragmentation can be seen. 
On the other hand,
in AMD-QL with a sufficiently large fluctuation strength $g_0$,
various fragmentations occur dynamically and the mass spectrum
in the dynamical stage is closer to the final post-Cascade results
as $g_0$ is increased. 
Considering the lifetime of the compound nucleus $\nuc{13}{N}^*$
and the predominance of pre\-equilibrium fragmentation
(which is evident from the forward-peaked angular distribution
of the fragments,
as is apparent in the 
data for the \nuc{9}{Be}/\nuc{11}{B} + p reactions at 45 MeV,\cite{p+B11}
for example), 
it appears that this new framework is superior to AMD, 
since the incorporation of the quantum fluctuation of the energy
leads to a significantly improved description
of  particle evaporation and fragmentation in dynamical process.
AMD-QL can describe the fragmentation of the moderately excited 
nuclei dynamically.
This feature is expected to be especially important
for the $\Xi^{-}$ absorption reaction,
because the total excitation energy in this reaction is very small
and a large part of stochastic two-body collisions are blocked.
The quantum fluctuations should then be the primary source in the system
for generating the fluctuations leading towards the various channels
in the final state.

We have employed AMD-QL for studying the dynamical mechanisms
for $\Lambda$ emission and twin hyperfragment formation
in the $\Xi^{-}$ absorption reaction.  
While AMD (plus Cascade) gives large production probabilities for 
   double hypernuclei (18\%)
   and cannot describe twin hyperfragment formation
   in the dynamical stage,
AMD-QL (plus Cascade) suppresses
   the formation of double hypernuclei ($\sim$10\%)
   and achieves a qualitative
   description of twin hyperfragment formation in the dynamical stage
   ($\sim$0.9\% in dynamical stage, and around 1.2\% after Cascade).
The calculated probabilities of double, single, and twin hyperfragment
formation are in the region allowed by the experimental estimate
at 90\% confidence level shown in Table~\ref{table:ULLimit}.
%
%
Moreover, the calculated single hyperfragment formation probability 
is within rough estimate, 26-73\%.
In addition, the double hyperfragment formation probability
becomes very close to the rough estimate, 3-9\%,
when the fluctuation strength is sufficiently large.
This is mainly due to the enhanced $\Lambda$ emissions.
In AMD, even when a $\Lambda$ has sufficient energy to escape 
the compound nucleus, it easily loses its energy by two-body
collisions before getting out.
Therefore, in AMD a $\Lambda$ can only escape from the nucleus
when it is created on the nuclear surface 
{\em and} its initial momentum is directed outwards.
By contrast,
in AMD-QL the quantum fluctuation continually kicks all the particles
and a $\Lambda$ can then be emitted even if it is initially bound.
Through these analyses,
we conclude that 
enhanced $\Lambda$ emission in the preequilibrium stage
is essential in describing the bulk dynamics of $\Xi^-$ absorption
at rest
and the quantum fluctuation given by the form of Eq.~\raf{eq:gmatform}
describes this enhancement well 
with the overall quantum fluctuation strength 
parameter $g_0$ being around 0.5.

\begin{figure}
\begin{center}
\geteps{12cm}{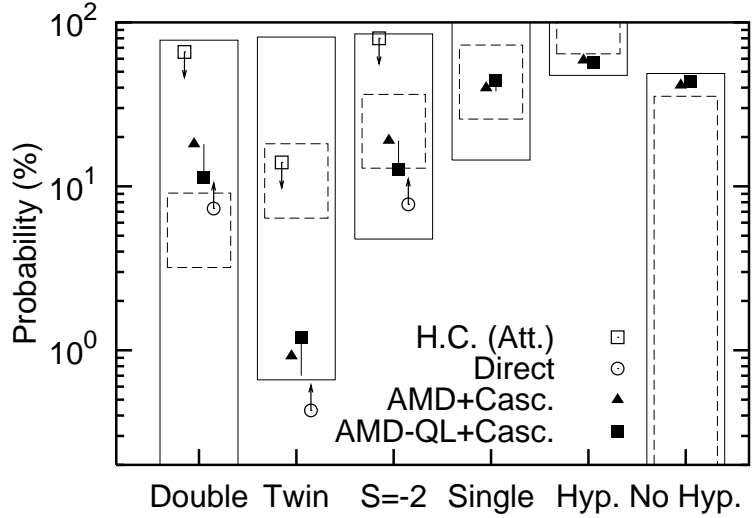}
\end{center}
\caption{%
Various model analyses of relative abundance of the various event types:
double, twin, $S=-2$, single, $S \leq -1$, and no hyperfragment formation.
Solid triangles and solid squares show the results calculated
with AMD and AMD-QL ($g_0=0.5$), and the error bars in AMD-QL
arise from the fluctuation strength parameter, from $g_0=$0 (AMD) to $g_0$=0.5.
The calculated results based on the double-hyperon compound nucleus
picture of $\nucll{13}{B}^*$ (H.C.) with attractive $\Lambda$-$\Lambda$
interaction~\protect\cite{Compound}
and the direct reaction picture for $\Xi^-$ absorption from $2p$ atomic
orbit~\protect\cite{YI97}
are shown by open squares and open circles, respectively.
The arrows from H.C. and direct reaction results
show the model uncertainty according to
the $\Lambda$ emission in the preequilibrium stage
and multistep processes other than direct break-up, respectively.
Experimentally allowed ranges are closed by solid lines
(90\% confidence level)
and dashed lines (rough estimate).
}
\label{fig:Limit}
\end{figure}

Although the calculated twin hyperfragment formation probabilities
are larger than the experimental lower limit (0.66\%),
it is still much below the rough estimate, 6--18\%.
Furthermore,
both of the two twin hypernuclear formation events
are in the channel \nucl{9}{Be}+\nucl{4}{H} 
and it is known that these twin hypernucleus are
formed after $\Xi^-$ is absorbed in \nuc{12}{C}.\cite{TwinA,TwinB} 
Therefore
the probability for finding the twin hyperfragments \nucl{9}{Be}+\nucl{4}{H} 
would reach around 13\%
(=2 events/ (31.1 events (C,N,O) $\times$ 0.48 (C)))
when the $\Xi^-$ is absorbed in \nuc{12}{C}.
The calculated probabilities for this channel are
only 0.16\% and 0.04\%, with AMD and AMD-QL ($g_0$=0.5), 
followed by Cascade, 
respectively. 
Even if we sum up the probabilities in the channels
\nucl{5}{He}+\nucl{4}{H}+\alp\ (0.12\% and 0.41\%) 
and \nucl{4}{H}+\lam+(2\alp,\nuc{8}{Be}) (0.09\% and 0.48\%)
to compensate the underestimate of the separation energy in \nucl{9}{Be},
the total probabilities, 0.37\% and 0.93\%, 
are still smaller than the above rough estimate by an order of magnitude.
%
%

%
%
This underestimate of twin hyperfragment formation
is not a specific problem in transport models,
and it was already pointed out in previous works.
%
%
As pointed out by Yamamoto, Sano, and Wakai,\cite{Compound}\ 
the statistical multifragmentation decay model of the double-hyperon
compound nucleus also fails to reproduce the experimental fact that 
the twin hyperfragment formation is more frequent
than that of double,
assuming that the $\Lambda$-$\Lambda$ interaction is attractive.
%
%
Yamada and Ikeda\cite{YI97} also showed that the channels of twin hypernuclei
gave very small spectroscopic factors,
and the calculated formation probability of twin hypernuclei was very small.

Figure~\ref{fig:Limit},
summarizes the results of various model analyses.
Since the emission probability of
one $\Lambda$ at the preequilibrium stage is not taken into account
in Ref.~\citen{Compound}, we assume that the calculated values
give the upper limit of the model.
On the other hand, the probabilities shown in Ref.~\citen{YI97}
for double, twin, and $S=-2$ in the direct reaction picture 
are considered to be lower limits, since the dissipation of
the doorway state into compound states and formation from these
compound states are neglected.
We can see immediately that all the analyses underestimate 
the twin hyperfragment formation probability
relative to that for double hyperfragment formation.

Therefore, if the twin hypernuclear formation probability, 
especially in the channel \nucl{9}{Be}+\nucl{4}{H}, is really high
as suggested, there must be some production mechanism
which is missed in all the theoretical works, including this work.
Another possibility is that the twin hypernuclear formation probability
is very close to or below the lower limit at 90\% confidence level
estimated by using the number of observed events.
In order to distinguish these two possibilities,
higher statistics experiments are required.
In addition to the higher statistics, 
one of the key challenges 
is to narrow down the upper limit in each type of events.
This requires resolving the unspecified events,
such as those events of type (C) in Table~\ref{table:Nakazawa}.

%
%
We have noted the shortcomings of the standard 
microscopic AMD model 
with which it is hard to describe
production of fragments near their ground states
in low-energy preequilibrium dynamical processes,
such as twin hypernuclear formation.
As the inclusion of quantum fluctuations leads to a significant improvement,
the extended model may yield a better description of other fission processes as well.

It is obviously important to broaden the confrontation of the theory
with experiment through application to other reactions for which
suitable data is available.
Moreover, it should be kept in mind that 
there are still ambiguities in AMD-QL.
%
%
The key quantity concerning fluctuations,
   the matrix $\bold{g}$, is determined phenomenologically.
Therefore, in the future, it would be of interest to determine $\bold{g}$
   from more fundamental considerations.
In this context,
we draw attention to a recently developed
extended AMD (called AMD-V) that incorporates a 
   diffusion of single-particle wave functions.\cite{OH96c} \ 
Since this description also introduces fluctuations
of the centroid parameter $\{\z_i\}$,
it may be helpful in determining
the matrix $\bold{g}$ of the AMD-QL model.
Another possibility worth exploring is 
to incorporate the effects of two-particle correlations
and their propagation~\cite{TDDM} in a statistical manner.
We hope that these challenges will help to deepen our microscopic understanding
of fluctuation and dissipation phenomena in nuclear dynamics.

\section*{Acknowledgements}

The authors thank Prof.\ S.~Shinmura, Prof.\ K.~Kat{\=o}, 
   Prof.\ K.~Imai, Dr.\ I.~Fuse, Prof.\ M.~Oka, Prof.\ K.~Matsuyanagi, 
   Prof.\ K.\ Nakazawa, and Dr.\ A.\ Ono
   for valuable discussions and suggestions,
   and all the members of the Nuclear Theory Group in 
   Hokkaido University for their great encouragements.
This work was supported in part by
the Grant-in-Aid for Scientific Research
(Nos.\ 07640365, 08239104 and 09640329)
from the Ministry of Education, Science and Culture, Japan,
and by the Director, Office of Energy Research,
Office of High Energy and Nuclear Physics,
Nuclear Physics Division of the U.S. Department of Energy
under Contract No.\ DE-AC03-76SF00098.



\begin{thebibliography}{99}

\bibitem{DoubleA}
M. Danyz {\it et~al.},
	Nucl. Phys. {\bf 49} (1963),  121.

\bibitem{DoubleB}
D.~J. Prowse,
	Phys. Rev. Lett. {\bf 17} (1966),  782.

\bibitem{DoubleC}
S. Aoki {\it et~al.},
	Prog. Theor. Phys. {\bf 85} (1991),  1287.


\bibitem{Multi}
N. Itoh,
	Prog. Theor. Phys. {\bf 44} (1970), 291;\\
N.~K. Glendenning,
	Phys. Lett. {\bf B114} (1982), 392;\\
G.~E. Brown, C. Lee, M. Rho, and V. Thorsson,
	Nucl. Phys. {\bf A567} (1994), 937;\\
J. Schaffner and I.~N. Mishustin, 
	Phys. Rev. {\bf C53} (1996), 1416.

\bibitem{BNL-E885}
M. May and G.~B. Franklin, AGS proposal E885.

\bibitem{BNL-E906}
T. Fukuda and R.~E. Chrien, AGS proposal E906 (1994).

\bibitem{KEK-E373}
K. Nakazawa, Nucl. Phys. {\bf A585} (1995),  75c.

\bibitem{TwinA}
S. Aoki {\it et~al.}, Prog. Theor. Phys. {\bf 89} (1993),  493.

\bibitem{TwinB}
S. Aoki {\it et~al.}, Phys. Lett. {\bf B355} (1995),  45.

\bibitem{Stopped}
K. Nakazawa,  in {\em Proc. of the 23rd INS International Symposium on Nuclear
  and Particle Physics with Meson Beams in the 1 GeV/c Region, Tokyo, Japan,
  Mar. 15-18, 1995}, edited by S. Sugimoto and O. Hashimoto (Universal Academy
  Press, Inc., Tokyo, Japan, 1995), p.\ 261.

\bibitem{Compound}
Y. Yamamoto, M. Sano, and M. Wakai,
	Prog. Theor. Phys. Suppl. {\bf 117} (1994), 265.

\bibitem{Yamada}
T. Yamada and K. Ikeda, Prog. Theor. Phys. Suppl {\bf 117} (1994),  445.

\bibitem{YI97}
T. Yamada and K. Ikeda, Phys. Rev. {\bf C56} (1997),  3216.

\bibitem{AMD}
A. Ono, H. Horiuchi, T. Maruyama, and A. Ohnishi,
	Prog. Theor. Phys. {\bf 87} (1992),  1185;
	Phys. Rev. Lett. {\bf 68} (1992),  2898;
	Phys. Rev. {\bf C47} (1993), 2652;
A. Ono, H. Horiuchi, and T. Maruyama,
	Phys. Rev. {\bf C48} (1993),  2946.

\bibitem{OR95}
A. Ohnishi and J. Randrup,
	Phys. Rev. Lett. {\bf 75} (1995),  596.

\bibitem{OR97a}
A. Ohnishi and J. Randrup,
	Ann. Phys. {\bf 253} (1997),  279.

\bibitem{OR97b}
A. Ohnishi and J. Randrup,
	Phys. Lett. {\bf B394} (1997),  260.

\bibitem{FMD}
H. Feldmeier,
	Nucl. Phys. {\bf A515} (1990),  147;\\
H. Feldmeier, K. Bieler, and J. Schnack,
	Nucl. Phys. {\bf A586} (1995),  493.

\bibitem{Cool1}
Y. Kanada-En'yo, H. Horiuchi, and A. Ono,
	Phys. Rev. {\bf C52} (1995),  628.

\bibitem{Cool2}
N. Itagaki, A. Ohnishi, and K. Kat{\=o},
	Prog. Theor. Phys. {\bf 94} (1995),  1019.

\bibitem{Lam-rang}
T. Motoba, H. Band\=o, K. Ikeda, and T. Yamada,
	Prog. Theor. Phys. Suppl. {\bf 81} (1985),  Chap. $\rm{3}$.

\bibitem{Lipkin}
H.~J. Lipkin, N. Mechkov, and A.~J. Glick,
Nucl. Phys. {\bf 62} (1965), 188; 199; 211.

\bibitem{Danielewicz}
P. Danielewicz and G.~F. Bertsch,
	Nucl. Phys. {\bf A533} (1991),  712.

\bibitem{NOH95}
Y. Nara, A. Ohnishi, and T. Harada,
	Phys. Lett. {\bf B346} (1995),  217.

\bibitem{SF96}
J. Schnack, H. Feldmeier, Nucl. Phys. {\bf A601} (1996), 181.
%

\bibitem{OH96b}
A. Ono and H. Horiuchi, Phys. Rev. {\bf C53} (1996),  2341.

\bibitem{OH96a}
A. Ono and H. Horiuchi, Phys. Rev. {\bf C53} (1996), 845.

\bibitem{SF97}
J. Schnack and H. Feldmeier, Phys. Lett. {\bf B409} (1997), 6.

\bibitem{OH96c} 
A. Ono and H. Horiuchi,
	Phys. Rev. {\bf C53} (1996), 2958.


\bibitem{OR93}
A. Ohnishi and J. Randrup,
	Nucl. Phys. {\bf A565} (1993),  474.

\bibitem{Risken}
H. Risken,
	{\em The Fokker-Planck Equation} (Springer, New York, 1989).

\bibitem{OR97c}
A. Ohnishi and J. Randrup,
	Phys. Rev. {\bf A55} (1997),  3315R.


\bibitem{Kiderlen}
D. Kiderlen and P. Danielewicz,
	Nucl. Phys. {\bf A620} (1997),  346.

\bibitem{Volkov}
A. Volkov,
	Nucl. Phys. {\bf 74} (1965),  33.

\bibitem{NijmegenDYN}
M.~M. Nagels, T.~A. Rijken, and J.~J. de~Swart,
	Phys. Rev. {\bf D15} (1977), 2547.

\bibitem{p+C12}
C.~T. Roche, R. G. Clark, G. J. Mathews, and V. E. Viola, Jr., 
	Phys. Rev. {\bf C14} (1976),  410.

\bibitem{Casc}
F. P{\"u}hlhofer,
	Nucl. Phys. {\bf A280} (1977),  267.

\bibitem{AMD2}
H. Takemoto, H. Horiuchi, and A. Ono,
	Phys. Rev. {\bf C57} (1998), 811.


\bibitem{p+B11}
R.~M. Devries, J.~W. Sunier, J.-L. Perrenoud, M. Singh,
G. Paic, and I. Slaus,
	Nucl. Phys. {\bf A178} (1972),  417.

\bibitem{Tadokoro}
S. Tadokoro, H. Kobayashi, and Y. Akaishi,
	Nucl. Phys. {\bf A585} (1995), 225c.


\bibitem{TDDM}
M. Gong and M. Tohyama, Z. Phys. {\bf A335} (1990),  153;\\
M. Gong, M. Tohyama, and J. Randrup, Z. Phys. {\bf A335} (1990),  331.

\end{thebibliography}
\end{document}